# MXene triggers high toughness, high strength and low hysteresis hydrogels for printed artificial tissue

Chendong Zhao[a, c], Yaxing Li[b, c], Qinglong He[a, c], Shangpeng Qin[a], Huiqi Xie[b,]* and Chuanfang Zhang[a,]*

**Affiliations:**

[a] College of Materials Science & Engineering, Sichuan University, Chengdu, Sichuan 610065 (P. R. China)

[b] Department of Orthopedic Surgery and Orthopedic Research Institute, Stem Cell and Tissue Engineering Research Center, State Key Laboratory of Biotherapy, West China Hospital, Sichuan University, Chengdu, Sichuan 610041 (P. R. China)

[c] These authors contributed equally to this work.

* Corresponding author: xiehuiqi@scu.edu.cn, chuanfang.zhang@scu.edu.cn.

**Abstract:** Substituting load-bearing tissues requires hydrogels with rapid processability, excellent mechanical strength and fatigue resistance. Conventional homogeneously polymerized hydrogels with short-chains/excessive branching exhibit low strength/toughness, being inadequate for artificial tissues. Here we introduce the heterogeneous polymerization-accelerated reaction kinetics on the $Ti_3C_2T_x$ MXene microreactor and sluggish kinetics beyond-to rapidly produce hydrogels within minutes. This allows the hyperbranched domains embedded within a highly entangled matrix, leading to excellent strength (2.4 MPa)/toughness (75.2 kJ m$^{-2}$) and low hysteresis (2.9%) in hydrogels superior to the rest ones. The rapid liquid-to-solid transition triggered by MXene suggests the great possibility of 3D printed robust hydrogels toward artificial tissue. Importantly, these printed hydrogels-based artificial ligaments have demonstrated impressive load-bearing capacity, wear resistance, and suturability compared to commercial analogs.

Tissue transplantation, while clinically indispensable for replacing dysfunctional tissues, still faces multiple challenges: critical donor shortages, immunological complications, and persistent ethical controversies (*1-3*). Emerging as a disruptive alternative, 3D-bioprinted hydrogel-based artificial organs leverage layer-by-layer-wise additive manufacturing to construct biomimetic architectures with anatomical precision (*4-6*). This paradigm shift not only circumvents the donor-dependency limitation but also enables biocompatible customization through patient-specific bioinks (*7, 8*). Current achievements demonstrate faithful replication of ventricle in vitro model (*9*) and vascularized alveolar model topologies (*10*), yet critical functional disparities still persist, particularly in dynamic mechanical performance (*11-13*). Current 3D-printed hydrogel systems exhibit failure under physiological stress regimes (tensile strength ＜500 kPa, fracture energy ＜ 100 J/m², hysteresis > 10%), being inadequate for tendon replacements or cardiac patches (*14-16*). To bridge this biomechanical gap, developing the expandable 3D printing strategy to manufacture high strength, high toughness hydrogels for the artificial organs is in urgent need.

3D printing is capable of rapid manufacturing of customized, prototyped hydrogels at any freedom (*17, 18*), opening vast opportunities in artificial organs such as blood vessels/hearts (*10, 19, 20*). Conventional hydrogels homogeneously polymerized in concentrated free radicals typically showcase compromised mechanical properties (*21-23*), due to abundant short molecular chains and rich side branches present in the hydrogel network. To this end, Xie et al elaborated precise molecular photopolymerization *via* introducing dynamic hindrance urea and side-chain carboxyl groups, leading to fast formation of hydrogel with increased toughness and mechanical strength (*24*). On the other hand, by segregating the rapid photoinitiated partial polymerization from the slow redox initiation process, Jason et al. achieved hydrogels with densely entangled long-polymer-chains and thereafter excellent toughness (*25*). To further break the bottleneck of rapid polymerization speed (allowing fast 3D printing) while preserving the hydrogels' mechanical strength/toughness, one needs to precisely manipulate the monomer crosslinking behavior such that the polymerization completes once the filaments depart from the nozzle. However, this has proven to be quite challenging.

Herein, we report on the heterogeneous polymerization strategy to revolutionize the hydrogel network enabled by two-dimensional titanium carbide MXene microreactor. Distinguished from the homogeneous initiator decomposition in traditional routes, the free-radical gradient induced by MXene accelerates the chain propagation rate of acrylic monomers along the microreactor, beyond which the polymerization kinetics is reduced. The former behaves as chemical cross-linking centers, generates hyperbranched chains and polymerizes into the prominent network, while the latter produces densely entangled chains, allowing 3D printing of hydrogels in seconds with adjustable strength (up to 2.35 MPa) and toughness (up to 75.2 kJ/mol), surpassing conventional casting hydrogels by 12~40 times in key mechanical metrics (shear/tensile/wear resistance). These 3D-printed, MXene-triggered hydrogels exhibit clinical-grade performance including exceptional mechanical stability (withstanding > 90 kg loads), surgical feasibility, and multi-fixation compatibility (screw/suture adaptability), while achieving precise anatomical restoration in porcine and human ligament/tendon reconstructions. These encouraging results indicate the unique role of MXene triggering chemistry in the heterogeneous polymerization, allowing direct rapid printing of biocompatible strong/tough hydrogels for multiple implanted artificial organs.

**MXene triggered heterogeneous polymerization**

Direct ink writing (DIW) allows the rapid printing of self-defined hydrogels (**Fig. 1**a). Traditional synthesis of hydrogels heavily relies on the homogeneous polymerization. The slow polymerization kinetics prolongs the liquid-solid transition time, being incompatible with DIW which requires the rapid prototyping after extrusion from the nozzle (*26-28*). To accelerates the kinetics, one needs to increase the free radical concentration of initiator by applying external energy (*29-31*). However, this potentially leads to rapid heat accumulation and burst polymerization, forming high-branching-degree polymer chains with a broad distribution of molecular weight (**Fig. 1b, i**). As such, conventional hydrogels typically exhibit inadequate mechanical strength or toughness for artificial organs such as ligaments, tendons, vascular conduits, and heart valve grafts. In other words, revolutionizing the crosslinking methodology of polymer chains such that strong, tough hydrogels can be directly 3D-printed is of significance to the artificial organs reconstruction and repairing.

Here we propose a MXene triggered heterogeneous polymerization strategy to accelerate the reaction kinetics as well as to break the bottleneck of mechanical properties in printed hydrogels (**Fig. 1b, ii and Supplementary Video S1**). The essence lies in the manipulation of chemical crosslinking behaviors enabled by the unique MXene triggering chemistry (**Fig. 1c**). Through wet-etching the $Ti_3AlC_2$ (**Fig. S1**) followed by delamination, well-defined edges and clean surface with pronounced (002) plane is observed in MXene flakes (**Fig. S2**). The persulfate groups spontaneously decompose to sulfate radicals once the former interacts with MXene (*32*). Theoretical calculation indicates that ammonium persulfate (APS) decomposes to sulfate radicals at room temperature (RT) spontaneously (the free energy, $\Delta G$= -5.14 eV), whereas nothing happens ($\Delta G$ raises to 0.71 eV) as MXene is absent (**Fig. 1d**). The electron paramagnetic resonance (EPR) detects the sulfate radical boom once MXene contacts APS (**Fig. S3**). Further calculations suggest that MXene sufficiently lowers down the barrier for monomer adsorption (**Fig. 1e**). The facts that both initiator decomposition kinetics and monomer adsorption are greatly accelerated on MXene flakes entitle MXene as the intrinsic microreactors, leading to much improved polymerization speed at RT. Thus, the rapid polymerization of monomers in the vicinity of MXene flakes is initiated and form hyperbranched regions, which serve as to improve polymer strength.

Beyond these microreactors, the polymerization speed dramatically reduces due to the radical concentration gradient, allowing the formation of highly entangled regions during the subsequent polymerization process. These entangled regions facilitate the release of stress during tensile deformation through sliding between molecular chains. The sharp difference of the polymerization speed across one liquid medium – the intrinsic definition of heterogeneous polymerization – indeed results in hydrogel with higher tensile strength (**Fig. 1f**), as well as surpassing toughness, modulus at a low hysteresis (**Fig. 1g**), compared to those of traditional casted hydrogels (TCH). The small-angle X-ray scattering (SAXS) results show an increased q value of MHPH in the range of 0.02 $Å^{-1}$ to 0.06 $Å^{-1}$, indicative of phase separation (**Fig. S4**). This is further confirmed by the modulus mapping derived from the atomic force microscopy (AFM), where high modulus (hyperbranched regions) and low modulus regions (highly entangled regions) are separated (**Fig. S5a**), in contrast to the uniform lower modulus in TCH (**Fig. S5b**). The unique phase separation endows the MXene triggered heterogeneous polymeric hydrogels (MHPH) with high toughness and tailored mechanical strength, as will be discussed below.

**High precision printing**

The rapid heterogeneous polymerization is examined by monitoring the rheological properties on a small-amplitude oscillatory shear machine, which minimizes the hindrance of shear force on polymerization (**Fig. 2a**). After adding few APS drops to the MXene-acrylate monomer ink, the whole solution behaves as liquid since the loss modulus (G'') dominates over the storage modulus (G') before ~42 s, beyond when G' exceeds G'', indicative of solidification (**Fig.2b**). Such a phase transition implies the critical switch of viscous fluid to elastic solid in a short time (*33*). When MXene is absent, the monomer solution showcases almost unchanged G' and G'' over 120 s, suggesting no phase transition and highlighting the importance of MXene in triggering the rapid heterogeneous polymerization. The solidification time is controllable by adjusting MXene amount at ease (**Fig. 2c**, **Fig. S6**), being significantly important for the DIW of robust hydrogels as one only needs to increase the MXene dosage to preserve the extruded filament shape. The fast temperature surge in the MXene-containing monomer solution is another direct proof of rapid heterogeneous polymerization, whose speed and solidification threshold can be quantitatively determined based on infrared thermal analysis (**Fig. S7**), agreeing well with the rheological tests (*34-36*).

While the onset solidification time governs the rapid prototyping, the swelling speed of extruded filaments largely affects the printing accuracy (*37, 38*). Here hydroxypropyl methylcellulose (HPMC) is used as the thickener to increase the spatial hindrance and to suppress the swelling of filaments through forming hydrogen-bonded crosslinked network (**Fig. 2d**). The HPMC content strongly influences the swelled volume (defined by integrating the contour changes, **Fig. 2e** and **Fig. S8**). For instance, the geometrical shape of extruded filament is well maintained by adding 4wt.% of HPMC even after 180 s (inset of **Fig. 2e**), in sharp contrast to the completely swelling filament as HPMC is absent. Such an anti-swelling capability is HPMC amount-dependent; the time required for a 10% swelled volume boosts from 5s in HPMC-free filament to 203 s in extruded filament containing 7 wt% HPMC (**Fig. 2e**). The much longer swelling time than the solidification time (42 s) ensures the efficient, high-accuracy 3D printing of filaments, and thereafter, ACLs. We note this is important for the implanted ACLs, since severe swelling generally leads to much-reduced mechanical strength in hydrogels (*39, 40*). Additionally, the thickener tends to form hydrogen bonding with MXene flakes (**Fig. S9**), leading to the uniform dispersion of MXene in the polymeric matrix (**Fig. S10**) which is conducive to the triggering chemistry of the latter.

To provide guidance for subsequent high-precision printing, **Fig. 2f** summarizes the effects of MXene and thickener contents on the solidification and swelling time (at a given 10% of volume change), respectively. When the solidification time is less than the critical swelling time, the ink can be smoothly extruded in high precision (**Fig. 2g ii**). Otherwise, issues such as unsolidification, fully swelling or low-accuracy printing may occur (**Fig. 2g**). Following these guidelines, a series of planar patterns and complex 3D structures are directly printed with high-precision (**Fig. S11a**), demonstrating excellent elasticity and resilience upon repeated compression and twist (**Fig. S11b** and **Supplementary Video S2**).

**Mechanical properties**

As mentioned above, the presence of concentrated sulfate free radicals and absorbed monomers on MXene microreactors allows the rapid polymerization and generates hyperbranched molecular chains. Beyond these MXene microreactors, diluted sulfate free radicals are slowly polymerized to form highly entangled molecular chains (**Fig. 3a**). Such a heterogeneous polymerization is time-dependent; longer soaking time develops the entangled chains in the printed hydrogels, thus increases the modulus (**Fig. S12**), tensile strength (**Fig. 3b**) and toughness (up to 75 kJ m$^{-2}$, **Fig. 3e i**) but decreases the elongation at the breaking point. Nevertheless, the strongest printed hydrogel still withstands 2300% of elongation (**Fig. 3b**). No lateral crack propagation is observed during the large tensile deformation (**Fig. 3c** and **Supplementary Video S3**). Since the MXene content affects the amount of hyperbranched regions (**Fig. 3d**), more MXene microreactors translate to improved tensile strength (**Fig. 3b and Fig. S12**) and tensile toughness till adding MXene up to 0.069% (**Fig. 3e, i**). This means that excessive hyperbranched regions and shortage of entangled regions deteriorate the anti-crack ability in printed hydrogels. The entangled density can also be affected by monomer/solvent ratio; a higher monomer concentration means more entangling regions, resulting in substantially increased modulus and toughness (**Fig. S13**). Furthermore, the hyperbranched topology improves the hydrogels' wear resistance and anti-swelling properties (*41, 42*), which are important parameters for organ implantation.

On top of excellent tensile strength/toughness, low hysteresis hydrogels allow the quick recovery of the pristine network without residual stress once relaxing the polymers. Importantly, our printed hydrogel demonstrates nearly ideal elasticity with minimal hysteresis even stretching by 400% (**Fig. 3f**). The hysteresis is down to 2.9% at 200% of stretching (**Fig. 3f, inset**), and can be well maintained upon repeated stretching (by 100%) for 100 cycles (**Fig. S14**). These printed hydrogels significantly stand out from the rest of double network or partially crystalline hydrogels, which typically exhibit either high toughness or low hysteresis, but hardly achieve both (**Fig. 3g** and **Table S1**). Considering the complication of implanting surgeries, our high toughness, low-hysteresis, stable hydrogels suggest the minimal invasion to human body without frequently replacing the parts. Indeed, these crucial parameters – mechanical strength, toughness and hysteresis– in MHPH have been overall achieved simultaneously, with values greatly outperformed those of reported hydrogels (**Fig. 3h**).

Despite ascorbic acid or ferrous sulfate also triggers the generation of free radicals for polymerization, however, the solidification time is two-order of magnitude more than that of MXene (**Fig. S15**), rendering the DIW impossible. The homogeneous formation of hyperbranching or highly entangled regions greatly reduces the tensile strength (by 30 times) and toughness (by 40 times) in conventionally triggered hydrogels (**Fig. S16 b,c**). This fully highlights the unique advantages of MXene microreactors-enabled heterogeneous polymerization in the rapid manufacturing of advanced hydrogels.

Indeed, MXene microreactors are also applicable for other monomers or thickeners to trigger the heterogeneous polymerization. Through simply tuning the ratio of acrylamide (AM) to solvent, the modulus and toughness of the printed PAM hydrogels can be accordingly adjusted (**Fig. 3i** and **Fig. S17**). As a result, MXene-triggered PAM sample exhibits a higher mechanical strength/modulus/toughness value than counterpart PAM samples reported in literature (**Fig. S18**). The printed PAM hydrogel exhibits a low hysteresis similarly as that of PAA one (**Fig. S19**). The heterogeneous polymerization of PAM or PAA can also be realized in poly (vinyl alcohol, PVA) thickener other from HPMC (**Fig. S20**), showing much higher mechanical strength and toughness in the printed PAA hydrogels (**Fig. 3i and Fig. S17**). These hydrogels with PVA as the thickener exhibit satisfactory adhesion performance to different substrates (**Fig. S21**), holding bright future in implanted tissues or biological patches. The freedom on the monomer and thickener choices suggests that the MXene microreactors-enabled heterogeneous polymerization to manufacture hydrogels with tailored strength/toughness and low-hysteresis, is universal. This immediately opens the door to 3D printed human organs such as vessels, muscles, heart valves and ligaments (**Fig. 3j**), where hydrogels with a variety of strength/toughness are preferred. Moreover, the printed MHPH shows no obvious toxic effects on cells (**Fig. S22**), as well as satisfactory anti-swelling properties after immersing in simulated body fluid (SBF) at 37 °C (**Fig. S23**), ensuring its promising applications in artificial tissues/organs that can be well maintained *in vivo*.

**Medical Application of MHPH as Artificial Ligament**

Prior to the artificial ligament/tendon reconstruction surgery, hydrogels should satisfy the strict mechanical performance upon fixation the screw and the bone tunnel (**Fig. 4a**). Impressively, the MHPH remains intact upon >2000 N of normal forces and withstands break forces equivalent to the weight of 90 kg human (**Fig. S24**). In addition, the mechanical strength of MHPH is hardly affected by the fixation using fastening screws, demonstrating excellent anti-shearing performance (**Fig. 4b**, **Fig. S25**), tensile strength (**Fig. 4c**) and wear resistance (**Fig. 4d** and **Fig. S26**), being 40, 12, and 30 times higher than those of TCH, respectively. The excellent resistance upon screw fixation laid the solid basis of soft and lightweight MXene-triggered hydrogel for potential ligament/tendon reconstruction.

Then, the 3D printed hydrogels (20 cm in length, 6 mm in width, and 0.5 mm in thickness, unless specified) are employed to reconstruct several clinical ligaments/tendons defects (**Fig. 4e**). When the ACL is ruptured, the digital radiography (DR) shows a significant anterior displacement of the tibia upon anterior drawer force, indicating a positive anterior drawer test (**Fig. S27a**). Such a test result turns negative again once reconstructed by our printed hydrogel, suggesting that the 3D printed hydrogel successfully restores the anterior stability of the knee (**Fig. S27a**). The knee with reconstructed ACL using two hydrogel strips easily withstand a fully filled bucket (total weight of 18.77 kg) suspended from the femur without pulling out and breaking (**Fig. S27b** and **Supplementary Video S4**), even when all knee ligaments and soft tissues were removed except the reconstructed ACL (**Supplementary Video S5**).

To gain more quantitative information of anterior translation of tibia in anterior drawer test, a single hydrogel strip was fixed for ACL reconstruction on the knee without medial femur condyle (**Fig. 4f**). By pulling the femur using 8 kg dumbbells (heavier than the clinically force strength, 6.8 kg), the horizontal distance between femur and tibia (DFT) in the reconstructed ACL dramatically reduces, similar to that of intact group but substantially lower than the ruptured one (**Fig. 4f** and **g**, and **Supplementary Video S6**). The calculated anterior translation of tibia ($AT_{tibia}$) value in the reconstructed ACL (1.3 ± 2.2 mm) is one order of magnitude lower than that of ruptured ACL (17.8 ± 1.9 mm), fully confirming the success of knee restoration enabled by the 3D printed hydrogel (**Supplementary method**).

Similar effects can be observed in the anterior talofibular ligament (ATFL) and calcaneofibular ligament (CFL) reconstruction by fixing a MHPH hydrogel strip on an ankle model (**Fig. 4h**). The varus stress test after reconstruction under DR demonstrates a talar tilt angle quite similar to that of intact state yet significantly lower than the rupture group (**Fig. 4i, j**). This agrees quite well with the trend of anterior translation of talus ($AT_{talus}$) under DR in **Fig. 4k and l**, where the reconstructed sample displays a substantially lower value than that of the ruptured group (**Fig. 4m**), suggesting the implanted hydrogel indeed restores the stability and function of ATFL and CFL. Indeed, the 3D printed hydrogel is also compatible with different clinical fixation methods to reconstruct ligaments/tendons. As demonstrated in **Fig. S28** and **Supplementary Video S6**, after the suture fixation of hydrogel strips on patellar ligament and anterior tibial tendon, respectively, the function of corresponding joints has been successfully restored. The resilient nature of the implanted hydrogel also facilitates the normal activities of the joint (**Fig. S29** and **Supplementary Video S7**).

Finally, to simulate a real clinical setting, we investigate the reconstruction of ATFL (**Fig. S30**) and Achilles tendon (**Fig. S31**) by inserting the 3D printed hydrogel strip into human amputated ankle. Real clinical reconstructive surgery with a MHPH hydrogel can be performed on a human limb at ease. The MSKUS measurement reveals a homogeneous echotexture of the hydrogel strip and similar ultrasound features on tendon and ligament, in good agreement with the homogeneous ligament/tendon-like signal under MRI, as shown in **Fig. S30d** and **S31b**. Furthermore, the presence of MHPH hydrogel does not interfere with the radiological detection of bony structures after reconstruction since it can't be developed under X-ray (**Fig. S27a**). These properties fit well with the requirements of imaging follow-up analysis for ligaments/tendons reconstruction, either monitoring surrounding bony changes or tracing materials *in vivo* changes, in the subsequent experiments and clinical applications. Therefore, our MXene-triggered hydrogel meets the requirements of mechanical properties, biocompatibility, fixation methods, surgical manipulations, and radiological detections for different ligament/tendon reconstruction, holding a bright future for clinical applications.

To summarize, we report on the rapid heterogeneous polymerization to achieved hydrogels with high tensile strength/toughness and low hysteresis. The essence of the heterogeneous polymerization lies in the unique MXene triggering chemistry and its crucial role of microreactors-the former greatly accelerates the decomposition kinetics of free radicals while the latter manipulates the monomer cross-linking behaviors through precisely adjusting the hyperbranched and/or entangled regions. Moreover, the critical rheological switch from liquid monomer dispersion to elastic solids triggered by the MXene allows the direct ink writing of strong yet tough hydrogels at any freedom. These 3D printed hydrogels have demonstrated bright potential in the reconstruction for complex mechanical and structures tissues/organs on clinic, such as artificial ligaments/tendons.

**Acknowledgements**

Dr. Chendong Zhao, Dr. Yaxing Li and Qinglong He contribute equally to this work. Prof. Cunjiang Yu and Prof. Yury Gogotsi are acknowledged for the suggestions. This work is generously supported by the Natural Science Foundation of China (grant no. 00301054A1073, 22479101, 22209118) and of Sichuan (grant no. 2024NSFSC0233, 2024NSFSC0002), the Fundamental Research Funds for the Central Universities (Nos. 1082204112A26, 20826044D3083, 20826041G4185, 20822041G4080) and Open research fund of State Key Laboratory of Mesoscience and Engineering (MESO-23-D06), the full-time postdoctoral research and development fund of West China Hospital of Sichuan University (2024HXBH076).

**Author contributions:**

 Conceptualization: *Chuanfang Zhang，Chendong Zhao，Yaxing Li，Qinglong He*

 Methodology: *Chuanfang Zhang，Huiqi Xie, Chendong Zhao，Yaxing Li，Qinglong He*

 Investigation: *Chendong Zhao，Yaxing Li，Qinglong He, Shangpeng Qin*

 Visualization: *Chendong Zhao，Yaxing Li，Qinglong He*

 Funding acquisition: *Chuanfang Zhang，Huiqi Xie*

 Project administration: *Chuanfang Zhang，Huiqi Xie*

 Supervision: *Chuanfang Zhang，Huiqi Xie*

 Writing – original draft: *Chendong Zhao，Yaxing Li，Qinglong He*

 Writing – review & editing: *Chuanfang Zhang，Huiqi Xie*

**Conflict of interest**

The authors declare no conflict of interest.

**Data availability**

**Supplementary Materials**

Materials and Methods

Supplementary Text

Figs. S1 to S30

Tables S1

References (*1–13*)

Movies S1 to S7

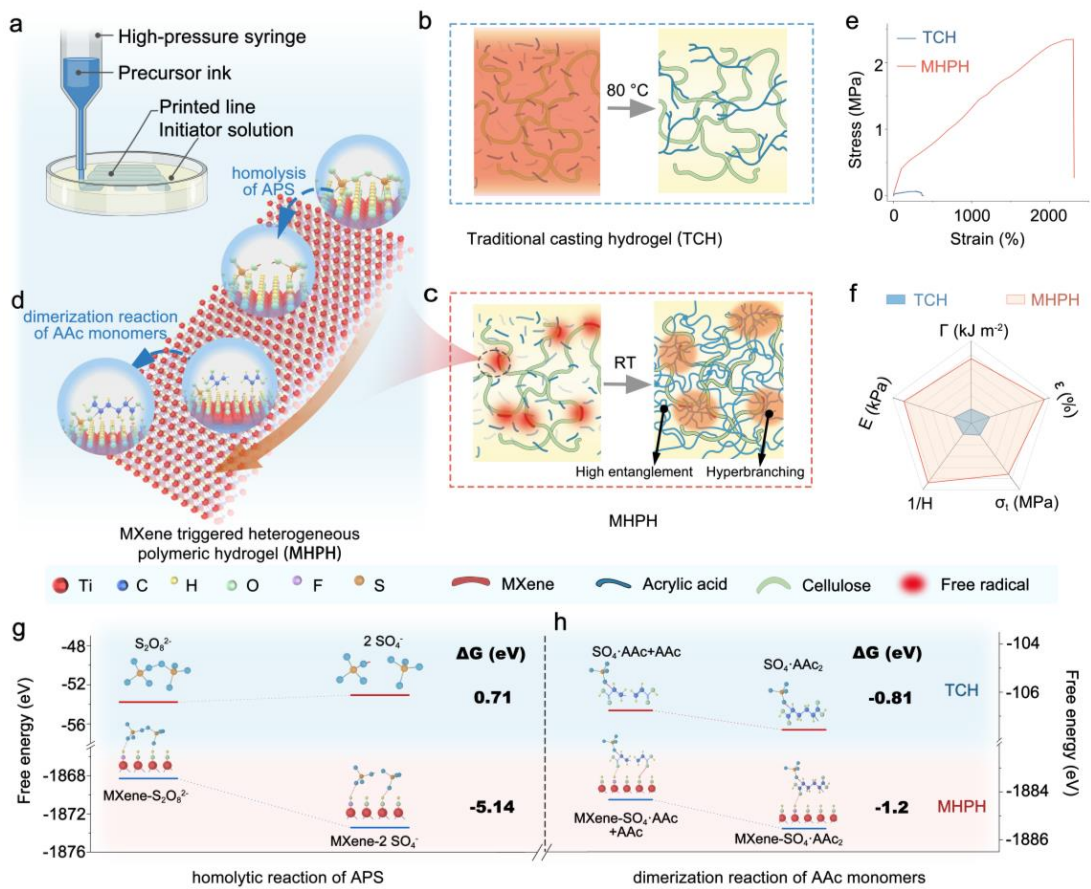

**Fig. 1. Preparation of high strength and high toughness hydrogel. a**, Scheme of 3D printing of MHPH. **b**, Schematic illustration of traditional casting hydrogel (TCH) at 80 ˚C. **c**, Mechanism of MXene trigger chemistry to synthesis MHPH at RT, where MXene behaves as microreactor to assist the rapid homolysis of initiator. **d**, Schematic of MXene as a microreactor triggering homolysis of APS and subsequent AAc polymerization. **e**, The tensile stress-strain curves and **f**, Comparison of modulus (E), toughness (Γ), strain (ε), tensile strength ($\sigma_t$) and 1/hysteresis (1/H) for MHPH and TCH. **g**, Comparison of free energy diagrams of APS decomposition and **h**, ΔG of AAc polymerization reaction routes with or w/o MXene.

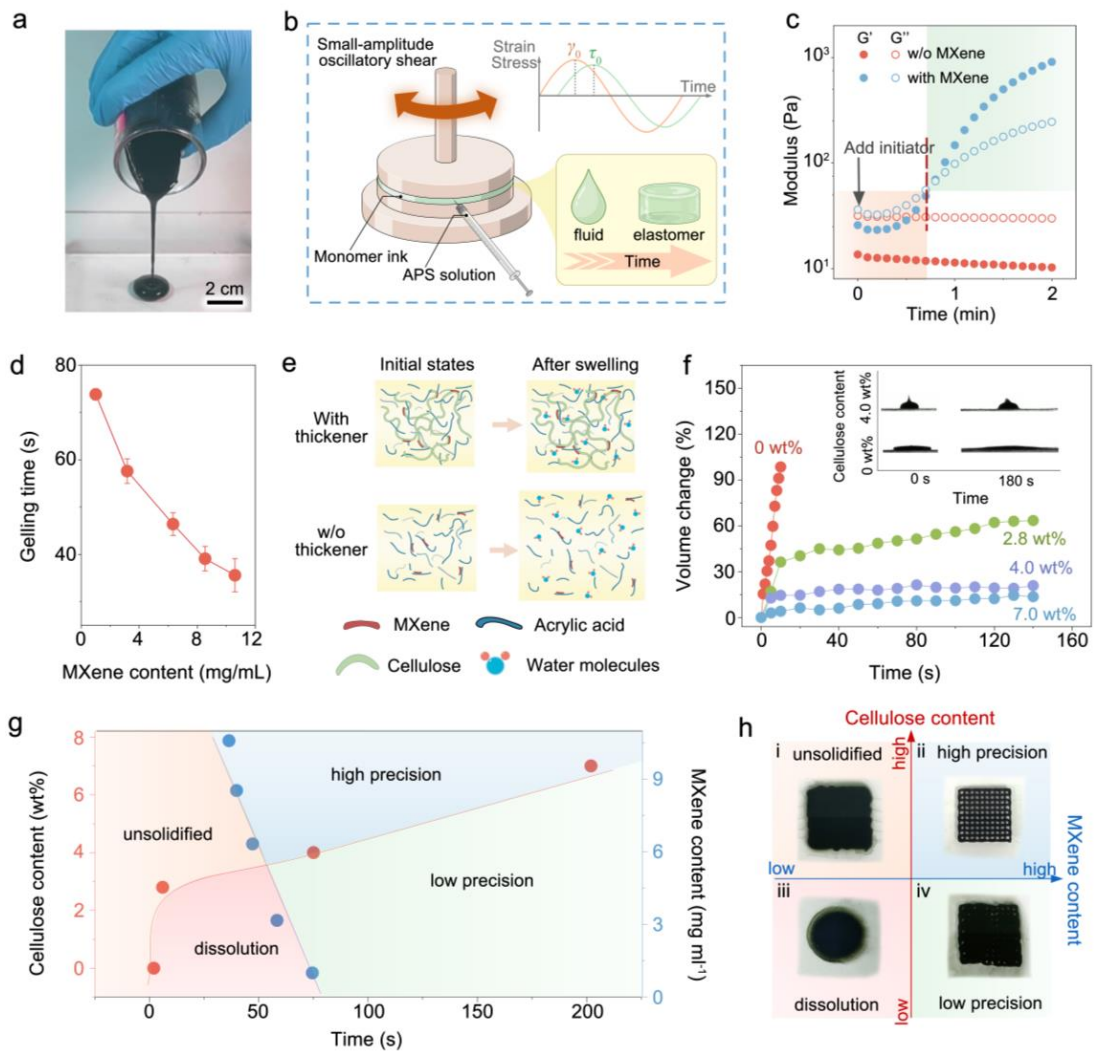

**Fig. 2. MXene assisted rapid gelation for extrusion printing. a**, Optical image of the MXene-monomer ink, showing its viscous nature. **b**, Schematic diagram of parallel plate rheometer testing. Inset shows the transition of liquid inks to elastic solid. **c**, The ink storage modulus (G') and loss modulus (G'') with or w/o MXene plotted as a function of time. 20 µL of APS solution is injected into the ink at 0 s. The dotted line shows the liquid to solid transition. **d**, Ink gelation time plotted as a function of MXene content. **e**, Diagrams of inks swelling with or w/o thickener in water. **f**, The volume change of printed filaments based on inks with different cellulose content over time. The inset shows the swelling of as-extruded filaments with 0 wt% and 4 wt% cellulose, respectively after 180 s. **g**, The guideline for high-precision 3D printing of MHPH. This map recommends reasonable cellulose and MXene contents such that stable patterns with well-defined structures can be obtained. Beyond the optimal cellulose and MXene contents, the printing is unsuccessful, as shown in **h**.

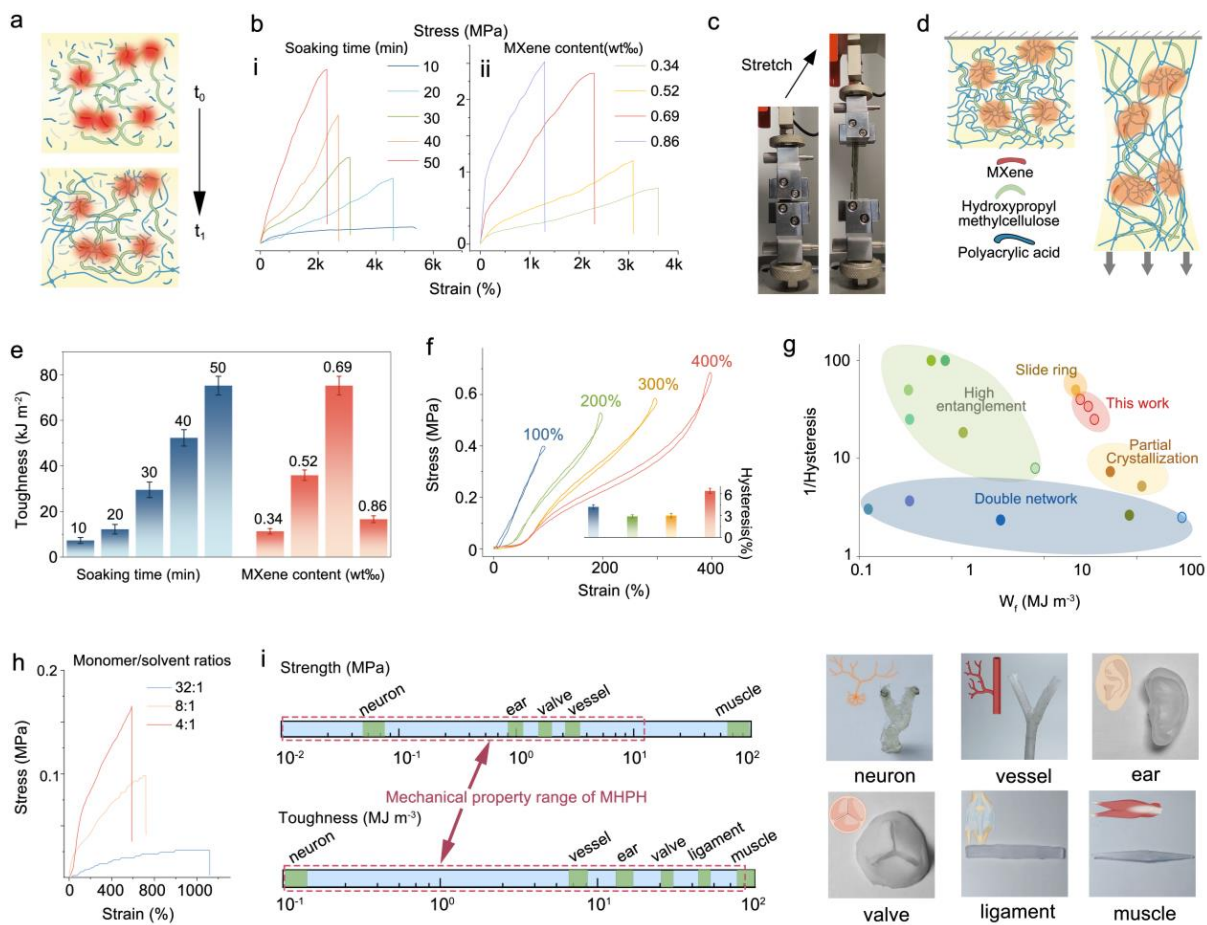

**Fig. 3. Mechanical performance of the printed hydrogel. a**, Schematic illustration of hydrogel cross-linking evolving with reaction time. Effects of **i** soaking time and **ii** MXene dosage on the tensile stress-strain curves, **b,** and toughness, **e,** of hydrogels after reaction for 50 minutes. **c**, The uniaxial tension of MHPH by 1000%. No cracks on the MHPH are observed upon stretching. **d**, The scheme of enhanced tensile strength enabled by MXene-triggered hyperbranched regions. **f,** Loading-unloading tensile stress-strain curves of hydrogels with different tensile degrees (soaking time is 50 min and MXene content is 0.69 wt‰. Inset shows the corresponding hysteresis statistics. **g**, Performance comparison of state-of-the-art hydrogels, showing that MHPH exhibits both high toughness and low hysteresis. **h**, The tensile curves of hydrogels with acrylamide as a monomer at different monomer/solvent ratios, showing the generality of MXene triggering chemistry in the rapid production of hydrogels with controllable mechanical properties on the MXene microreactors. **i**, The strength-toughness coverage of as-printed hydrogels for artificial human organs, showing broad promising applications in the organ transplantation using our MXene-triggered printed hydrogels. Also included are some indispensable organ models.

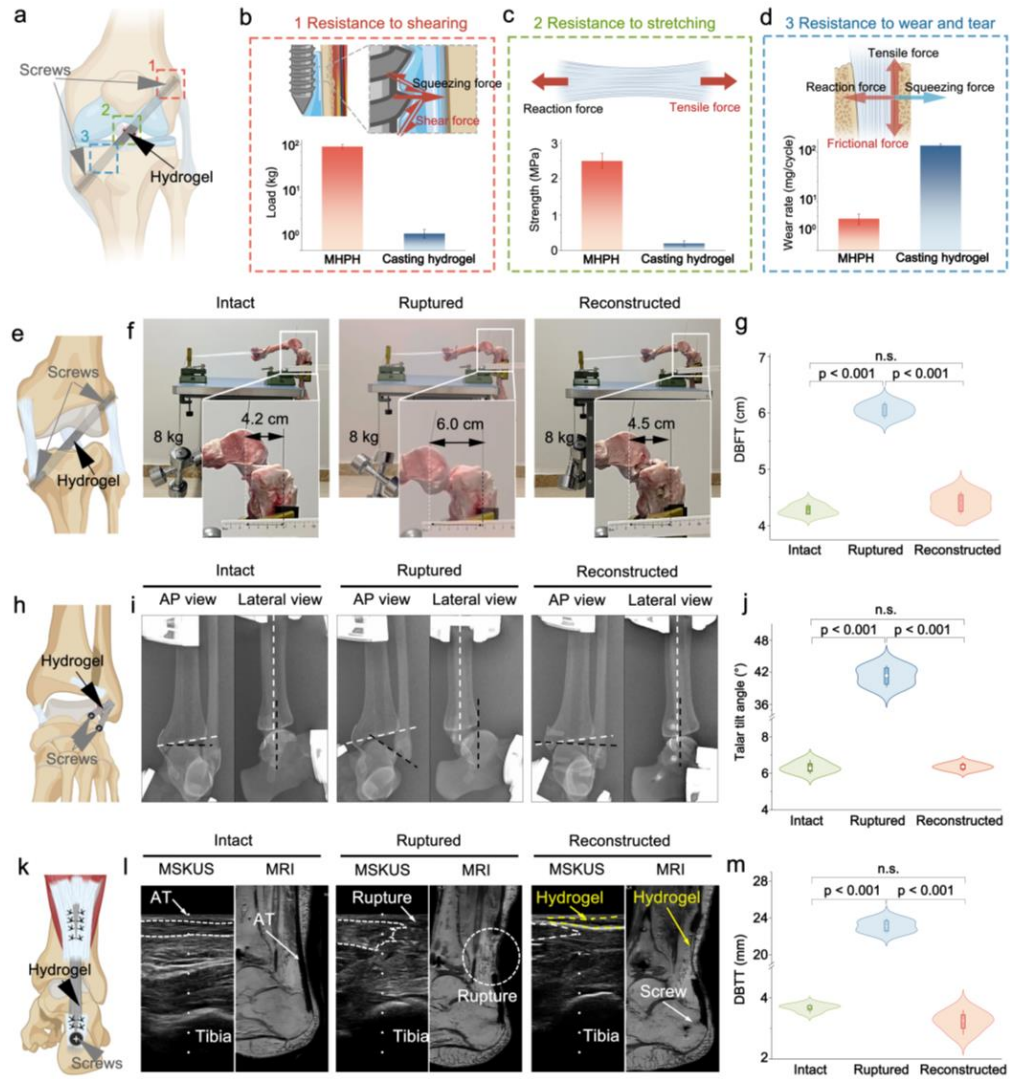

**Fig. 4. Applications of MHPH for ACL reconstruction, anterior talofibular ligament (ATFL) reconstruction and calcaneofibular ligament (CFL) replacement. a**, Schematic diagram of MHPH for ligament surgery. Comparison of **b**, cutting resistance (The load that can be sustained without being damaged by screws), **c**, tensile strength and **d**, wear resistance performance between MHPH and traditional casting hydrogel. **e**, Schematic diagram MHPH for ACL reconstruction. **f**, Quantitative displacement testing and **g**, corresponding numerical statistics of knee joint stability before and after reconstruction. **h**, Schematic diagram of the ATFL reconstruction surgery for human ankle. **i**, The digital radiography (DR) of the ATFL replacement. The **j**, angle and **m**, displacement numerical statistics of the replaced ankle joint. **k**, Schematic diagram of MHPH for CFL reconstruction. **l**, The musculoskeletal ultrasound (MSKUS) and magnetic resonance imaging (MRI) of the reconstructed CFL.

# Supplementary Materials for

**MXene triggers high toughness, high strength and low hysteresis hydrogels for printed artificial tissue**

*Chendong Zhao[a, c], Yaxing Li[b, c], Qinglong He[a, c], Shangpeng Qin[a], Huiqi Xie[b,] * and Chuanfang Zhang[a,] **

[a] College of Materials Science & Engineering, Sichuan University, Chengdu, Sichuan 610065 (P. R. China)

[b] Department of Orthopedic Surgery and Orthopedic Research Institute, Stem Cell and Tissue Engineering Research Center, State Key Laboratory of Biotherapy, West China Hospital, Sichuan University, Chengdu, Sichuan 610041 (P. R. China)

[c] These authors contributed equally to this work.

* Corresponding author: xiehuiqi@scu.edu.cn, chuanfang.zhang@scu.edu.cn.

**The PDF file includes:**

    Materials and Methods
    Supplementary Text
    Figs. S1 to S30
    Tables S1
    References

**Other Supplementary Materials for this manuscript include the following:**

    Movies S1 to S7

**Materials**

Titanium powder (99.5%), aluminum powder (99.5%), graphite (99%), acrylic acid (AAc, AR), hydroxypropylmethyl cellulose (HPMC, 15000 mPa.s), N, N'-methylene diacrylamide (MBAA, 99%), acrylamide (AM, AR), polyvinyl alcohol (PVA, 1788), ascorbic acid (AA, AR), iron sulfate heptahydrate (99.9%), N,N-Dimethylformamide (DMF, 99.9%), ammonium persulphate (APS, 98%, Adamas), polyurethane (PU, 500 mesh, Guangyuan Plastic Products Co., Ltd, Guangzhou) were used as purchased without further purification.

**Methods**

**Synthesis of the $Ti_3AlC_2$ MAX Phase**

Graphite, titanium powder and aluminum powder were mixed and ball milled for 18 hours in a molar ratio of 1.8:3:2.2. Then the mixed powder was heated to 1600 °C and held for 8 hours in an argon atmosphere. After natural cooling, the sintered brick was crushed and sieved through a 400 mesh sieve to obtain particle sizes below 38 μm. Next, the powder was stirred with 9 M hydrochloric acid at 40 °C for 72 hours to remove impurities of intermetallic compounds. Finally, the powder was completely rinsed with deionized for multiple times and vacuum-dried at 80 °C for 6 hours. The as-obtained MAX powder was stored in the glove box for further use.

**Wet-etching of the $Ti_3C_2T_x$ MXene**

The synthesis of $Ti_3C_2T_x$ MXene was performed following the recipe of ref (*1*). One gram of $Ti_3AlC_2$ MAX powder was slowly added to 20 mL of the mixture of concentrated HF (49 wt %), concentrated HCl (36 wt %), and DI-$H_2O$ with a volumetric ratio of 1:6:3. The MAX phase was etched at 35 °C for 48 hours. Then, MXene was washed with DI water for multiple cycles of centrifugation until the pH reached ~7.

**Delamination of $Ti_3C_2T_x$ MXene**

To intercalate $Li^+$ ions between the MXene layers, the multilayered MXene was stirred at 300 rpm in a LiCl solution (the solution concentration of 20 mg/mL) at 35 °C for 18 h. Then, LiCl was removed by repeatedly centrifuging for 10 min at 3500 rpm, discarding the upper liquid and redispersing the bottom mud in DI-water for 8 times. To collect single-layer $Ti_3C_2T_x$ nanosheets, the bottom centrifuged MXene mud were redispersed in DI-water and vigorously shaken by a vortex machine for 30 min. Subsequently, the mixed solution was centrifuged at 3500 rpm for 1 hour to obtain the upper dispersion consisted of few-layered MXene, which was further centrifuged at 12000 rpm for 1 hour. The bottom centrifuged MXene mud was further homogenized using a Mayer rod and added few drops of water to obtain MXene inks with desired concentration. To calculate the solid content C of obtained MXene ink, M grams of ink was taken and vacuum-dried at 80 °C for 12 hours, then the dried powder was weighed as m grams. After repeated the above steps for ten times, the solid content C can be calculated as C = m/M.

**Characterization of MXene**

The structure of MXene was confirmed by XRD (Cu Kα radiation, λ = 1.54178 Å, D/max 2550 V, Rigaku) and Raman spectroscopy (Horiba LabRAM HR Evolution), the morphology was studied by SEM (ZEISS Sigma 300) and TEM (Japan-JEOL-JEM 2100Plus), the elemental composition was characterized by EDS (OXFORD X-act one) and XPS (Thermo Scientific ESCALAB Xi+).

**Extrusion printing**

**Ink preparation:** The ink for extrusion-printing was consisted of acrylic acid as the monomer, MBAA as the crosslinking agent, hydroxypropyl methyl cellulose as the thickener and MXene as the trigger. Define W as the molar ratio of water to monomer and C as the molar ratio of crosslinking agent to monomer. Firstly, a small amount of thickener is mixed with water (the mass ratio of thickener to water is 1:50). The mortar grinder was used for ink homogenizing at a speed of 1500 rpm for 30 minutes, then different mass of MXene slurry was added and stirred for another 30 minutes until the slurry is completely mixed. Finally, acrylic acid, MBAA, and additional thickeners were added and continue stirring for 60 minutes, forming the homogenized, extrudable ink.

**Initiator solution preparation:** APS was mixed with water in a mass ratio of 1:16 and stirred for 30 min to obtain the initiator solution.

**3D printing:** The extrusion printing process for the fabrication of high-toughness hydrogel was performed using a programmable three-axis pneumatic robotic deposition system (SM500ΩX-3ASS, MUSASHI) equipped with painting software (MuCAD, MUSASHI). The prepared ink is extruded into the initiator solution through the cylindrical nozzle (14G) under a constant air pressure (80–150 kPa), moving speed (5–10 mm s$^{-1}$) and distance between needle and base (350–450 μm).

After extrusion printing, the hydrogels were taken out of the initiator solution at different interval time to achieve varied degrees of reaction. The as-printed hydrogels were rinsed with deionized water for 10 min to remove the superficial residual APS, followed by the annealing in an oven at 50 °C for 1 hour to complete the polymerization.

**Preparation of artificial ligaments:** Given the prolonged mechanical loading of artificial ligaments in physiological fluid environments and the swelling tendency of poly(acrylic acid) (PAA) in aqueous media, we substituted hydroxypropyl methylcellulose (HPMC) thickener with polyurethane (35 wt% of the ink mass). The artificial ligaments were fabricated using the same 3D printing parameters described above. This sample is only used in pig anterior cruciate ligament, human anterior talofibular ligament and Achilles tendon implantation.

**Rheological characterization**

Rheological characterization was carried out on a rotational rheometer (MCR302, Anton Paar GmbH, Austria). All steady-shear and dynamic rheological tests were performed utilizing a 20 mm stainless steel parallel plate at room temperature. Dynamic stress sweeps were performed at a constant angular frequency of 1 rad s$^{-1}$. Dynamic frequency sweeps were done at a constant strain amplitude within the linear viscoelastic regime of each sample.

**Tensile testing**

The uniaxial tensile tests were carried out on the electronic tensile machine (*2*) with a strain rate of 100 mm min$^{-1}$. A minimum of three specimens were tested for each group. Dogbone-shaped patterns were printed with a width of about 2 mm and a thickness of 1 mm (The accurate dimensions were measured by a vernier caliper) (ISO-37). The end of printed strip was clamped with a sandpaper to prevent the sliding or the stretching of shoulder. Consequently, no sliding was observed during the entire uniaxial tensile test, all samples fractured in the middle of the linear region. Stress and strain were calculated through dividing the measured force by the initial cross-sectional area and dividing the displacement by the initial distance between clamps. The modulus were obtained from the initial slope of the stress–strain curves.

**Pure shear tests**

Two identical rectangular hydrogel strips with 40 mm long, 20 mm wide and ~ 1 mm thick were 3D printed (one cut with an 8 mm notch located in the middle of the sample edge, another unnotched). The initial clamping distance for test is 3 mm. The critical strain $\varepsilon_c$ was obtained by uniaxial tension of notched specimens until the unstable crack propagation. And the stress-strain curve was obtained through the same tensile test for the unnotched specimen. Finally, integrating the obtained stress-strain curve from 0 to $\varepsilon_c$ and then multiplying it with the initial clamp distance (H) to obtain the fracture energy (Γ) (*3, 4*).

$$\Gamma = H \int_0^{\varepsilon_c} \sigma d\varepsilon$$

**Calculation method**

All density functional theory (DFT) calculations are performed in the Vienna ab initio simulation package (VASP) (*5*), which is based on a set of planar wave bases and employs the projection augmented wave method (*6*). The Generalized Gradient Approximation (GGA) and Perdew Burke Ernzerhof (PBE) parameterization were used to handle exchange-correlation potential (*7*). The van der Waals correction of Grimme's DFT-D3 model was also adopted (*8*). A supercell of MXene (0 0 2) surface containing 5 × 5 × 1 unit cells were used to adsorbate molecule. At the same time, a vacuum area of approximately 15 Å was applied to avoid interaction between adjacent images. The energy cutoff is set to 450 eV. The 1×1×1 k-points sampling with Gamma method were used. The structures were completely relaxed until the maximum force on each atom was less than 0.02 eV/Å, and the energy convergence standard was 10$^{-5}$ eV.

The adsorption energy ($E_{ad}$) is calculated by the following equation: $E_{ad} = E_{total} - E_{M1} - E_{M2}$, where $E_{total}$, $E_{M1}$ and $E_{M2}$ represent the calculated total energy, the energy of M1 molecules, and the energy of M2 molecules, respectively.

**Wear rate measurement**

The wear rate was measured on a home-made setup (9). The 1 cm-wide, 1 cm-long testing specimens were fixed on the test stand. Then a pressure of 16 N (160 kPa) was applied from the top and the 120 grit sandpapers were used as friction substrate. A reciprocating linear motion was performed at a speed of 200 mm/min and a stroke of 10 cm. The wear rates were calculated by weighing the material before and after wearing. The sandpaper was replaced for each sample.

**CCK8 method and Live/Dead staining**

CCK8 method and Live/Dead staining were used to detect the survival of NIH3T3 cells co cultured with samples after 12 h, 24 h, 36 h, 48 h, and 60 h. The hydrogel samples were sterilized with high-temperature and high-pressure steam (121 °C, 20 min) and were cooled to room temperature. Then the samples were added into the culture medium with a 10 mg/mL concentration and extracted for 24 hours in a constant temperature incubator with 5% $CO_2$ at 37 °C. Next the extraction mother liquor was obtained through suction filtration with the 0.22 μm membrane and diluted 100 times to 0.1 mg/mL with culture medium to obtain the working concentration.

**Ligament/Tendon Reconstruction**

The 3D printed hydrogels (size: 20 cm in length, 6 mm in width, and 0.5 mm in thickness, unless specified) were employed to reconstruct several common clinical ligaments/tendons failures.

For anterior cruciate ligament (ACL) reconstruction, the fresh knees of crossbred pigs were obtained from the supermarket. The surrounding muscle, patella, infrapatellar fat, and joint capsule were removed, while the collateral ligaments, medical and lateral meniscus, and both cruciate ligaments were retained. The ACL was cut off and exposed. Two bone tunnels were created within the ACL tibial and femoral footprints by a 5.0 mm-diameter twist drill. Two hydrogel strips (size: 20 cm in length, 6 mm in width, and 0.5 mm in thickness) were passed through the bone tunnels to replace the ACL and were fixed by two polyether ether ketone (PEEK) interface screws. Anterior and posterior drawer tests under digital radiography (DR) were performed with standardized clinical methods by an orthopedic surgeon with 10 years of experience. Then, to demonstrate the mechanical properties of the knee joint after the ACL reconstruction by hydrogel, and to evaluate whether the posterior cruciate and collateral ligaments remain intact or incomplete, the tibia was fixed and an 18.9 L bucket filled with water (18.77 kg) was suspended from the femur.

For better exposure of ACL, the medial femur condyle, medial collateral ligament, and posterior cruciate ligament were removed from the knee. The tibia was fixed on the experimental table, and the femur was pulled by dumbbells (total weight of 8 kg, which is more than the clinically used force strength of 6.8 kg) to measure the free anterior displacements of the tibia in relation to the femur. Two Kirschner wires were inserted in distal femur and tibial tuberosity, respectively, as reference marker location, and distance between femur and tibia (DFT) was measured as the horizontal distance between the insert points of the Kirschner wires in femur and tibia. The measurements of three group were sequentially performed (from intact group to ruptured group to reconstructed group) on the same sample, and the data of intact group was used as normal control. The anterior translation of tibia ($AT_{tibia}$) was calculated as: $AT_{tibia\ (ruptured/reconstructed)} = DFT_{(ruptured/reconstructed)} - DFT_{(intact)}$. $AT_{tibia} > 5$ mm was defined as positivity for anterior drawer test, conversely, was a negative result.

Then, biomechanical tests were performed using a tensile testing machine for the knee samples, in which the medial femur condyle was removed. The femur and tibia were immobilized firmly in clamps. Prior to the test, the specimen was preloaded with a preload of 1 N for 5 min. After preconditioning, the ultimate failure load was investigated with an elongation rate of 2 mm/min. The failure load was recorded from the load-time curve. The test was completed when the hydrogel was ruptured or was pulled out from the bone tunnel.

To better show the functional recovery of the ankle, a procedure of talofibular ligament (ATFL) and calcaneofibular ligament (CFL) by using the MHPH hydrogel was performed on the ankle model. Bone tunnels were made in the talar neck, distal fibula, and calcaneal body. A hydrogel strip was passed through the bone tunnels to replace the ATFL and CFL and was fixed by two PEEK interface screws. The activities of ankle joint after reconstruction were examined. Varus stress test and anterior drawer test under DR were performed with standardized clinical methods by an orthopedic surgeon with 10 years of experience to detect the stability of the ankle. The talar tilt angle and the free anterior displacements of the talus in relation to the tibia were measured. Distance between talus and tibia (DTT) was recorded as the horizontal distance between the anatomical talus axis and anatomical tibial axis. The measurements of three group were sequentially performed (from intact group to ruptured group to reconstructed group) on the same sample, and the data of intact group was used as baseline values to calculate the anterior translation of talus ($AT_{talus}$). Specifically, $AT_{talus}$ was calculated as: $AT_{talus\ (ruptured/reconstructed)} = DTT_{(ruptured/reconstructed)} - DTT_{(intact)}$. $AT_{talus} > 5$ mm was defined as positivity for anterior drawer test, conversely, was a negative result. Positivity for varus stress test was defined as talar tilt angle > 15 degree.

For anterior ATFL reconstruction, ankle models and a human specimen were used. The human ankle sample was obtained from amputated leg. The ATFL was exposed via a lateral incision. The ATFL has ruptured in this human ankle sample before amputation. Two bone tunnels were created in talar neck and distal fibula, respectively, by a 5.0 mm-diameter twist drill. A MHPH hydrogel strip (size: 10 cm in length, 6 mm in width, and 0.5 mm in thickness) was passed through the bone tunnels to reconstruct the ATFL and was fixed by two PEEK interface screws. The musculoskeletal ultrasound (MSUS) and magnetic resonance imaging (MRI) were performed for signal detection of the ligament and hydrogel, and a normal ankle from hospital staff was selected as normal control.

For patellar ligament reconstruction, the fresh lower limbs of crossbred pigs were obtained from a market. The skin and muscle were removed, and the patella and patellar ligament were exposed. A defect (2 cm in length) was made in the middle of the patellar ligament. A hydrogel strip (size: 4 cm in length, 6 mm in width, and 0.5 mm in thickness) was stitched to bridge both ends of the tendon defects with 0# non-absorbable sutures. The functional recovery of knee extension was examined.

For anterior tibial tendon reconstruction, the fresh lower limbs of crossbred pigs were obtained from the supermarket. The anterior tibial muscle was exposed, and the surrounding skin and other muscles were removed. A defect (2 cm in length) was made in proximal anterior tibial tendon. A hydrogel strip (size: 4 cm in length, 6 mm in width, and 0.5 mm in thickness) was stitched to bridge both ends of the tendon defects with 0# non-absorbable sutures. The tibia was fixed on the experimental table and a dumbbell (total weight of 4 kg) suspended on the feet. The functional recovery of ankle joint dorsiflexion driven by the reconstructed tendon was examined.

For Achilles tendon reconstruction, a human ankle sample was obtained from amputated leg. The Achilles tendon was exposed via a longitudinal incision and a defect (3 cm in length) was created. Two hydrogel strips (size: 10 cm in length, 6 mm in width, and 0.5 mm in thickness) were stitched to bridge both ends of the tendon defects with 0# non-absorbable sutures. The distal hydrogel was passed through a calcaneal bone tunnel and fixed by a PEEK interface screw. The MSUS and MRI were performed for signal detection of the ligament and hydrogel.

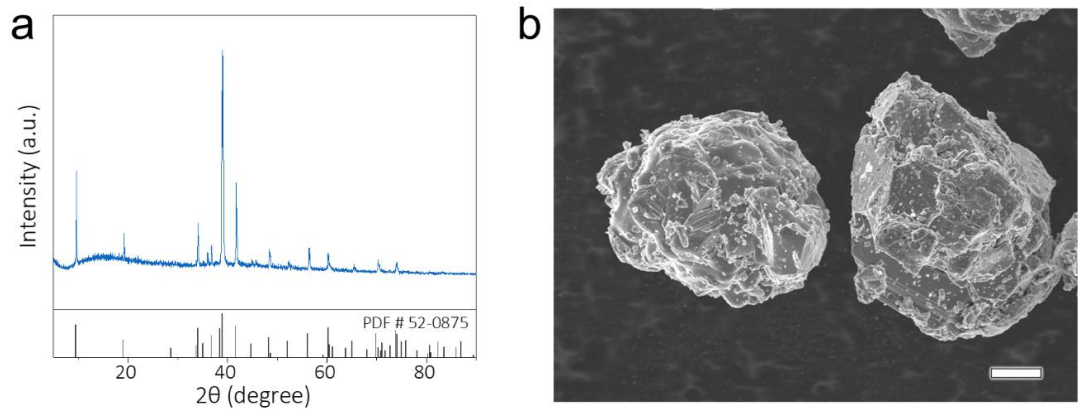

**Fig. S1. a**, The XRD and **b**, SEM image of the as prepared MAX phase, the scale bar is 5 μm. The XRD result indicates that the MAX phase is high purity $Ti_3AlC_2$, with all peaks well-indexed with the standard PDF card. The SEM image shows that the as-synthesized powder exhibits a ternary-layered morphology, with a grain size ~ 25 μm.

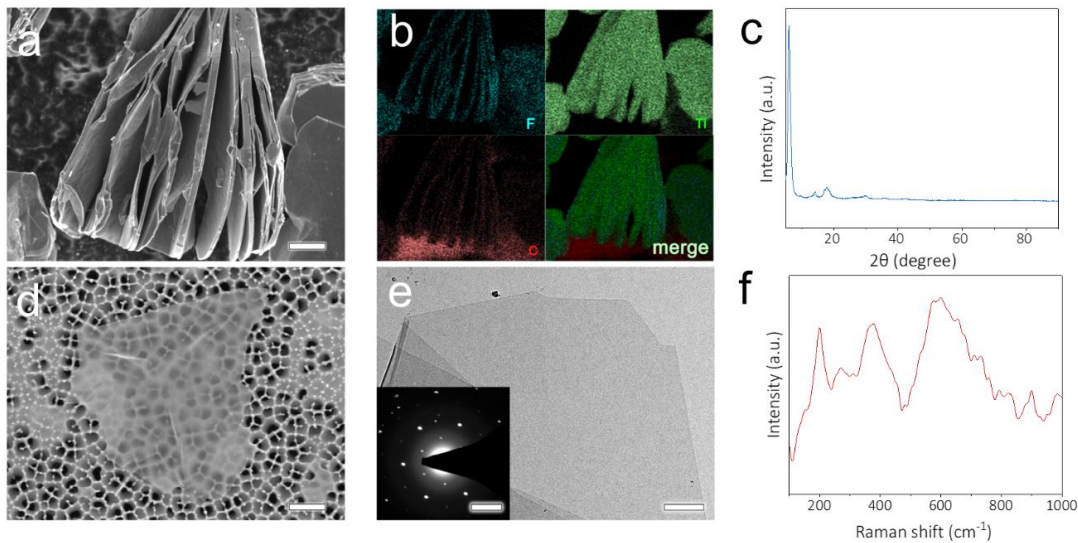

**Fig. S2**. Characterization of the as the prepared MXene nanosheets. **a**, Morphology, scale bar is 2 μm, **b**, EDS mapping and **c**, crystal phase characterization of multilayered MXene. SEM (**d**) and TEM (**e**) image of exfoliated MXene nanosheets, scale bar is 500 nm (inset in e is selected area electron diffraction, SAED, scale bar is 5 1/nm). **f**, Raman characterization of the delaminated single-layer MXene.

The minimally intensive layer delamination (MILD) method was used to etch the MAX phase. The etched product shows a typical accordion like structure. Relatively thick sheets are observed on the multilayered MXene solid, which is due to the slow reaction rate and gas release compared with the pure HF etching. However, the XRD results shows that the MAX phase has been fully etched as the (002) peak has shifted from the original 9.65° of MAX phase to around 6° without any impurities. The EDS mapping indicates the uniform distribution of elements in the etched crystal.

The delaminated MXene dispersion was dropped onto a porous anodized aluminum film for observation of the number of MXene layers through contrast. The TEM and SAED images suggest that the obtained MXene is a single layer with good crystallinity. the Raman results indicate there is no apparent oxidation during the etching and delamination processes, suggesting the high-quality nature of the as prepared MXene nanosheets.

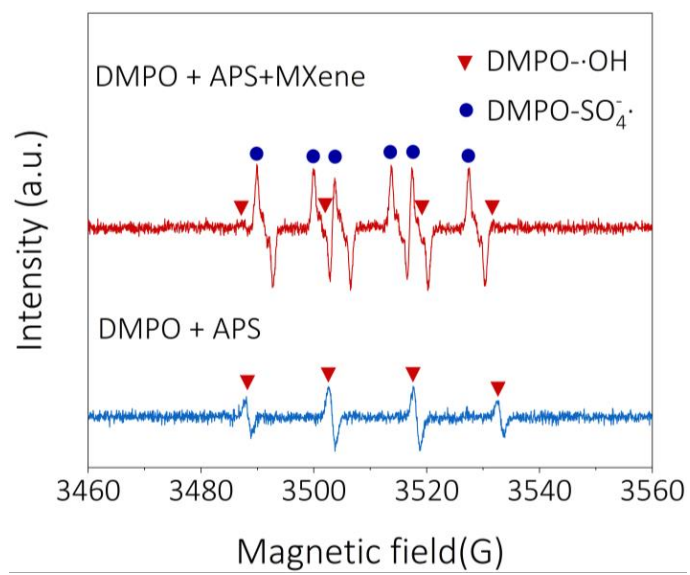

**Fig S3**. EPR spectra for the detection of radicals under the systems with or w/o MXene in the presence of DMPO.

Based on the EPR tests, it clearly shows that once the system containing MXene, apparent signals corresponding to sulfate radicals are observed. These sulfate radicals signals are so strong that they almost overwhelm the hydroxyl groups signals. It's thus fair to deduct that MXene triggers the production of sulfate radicals.

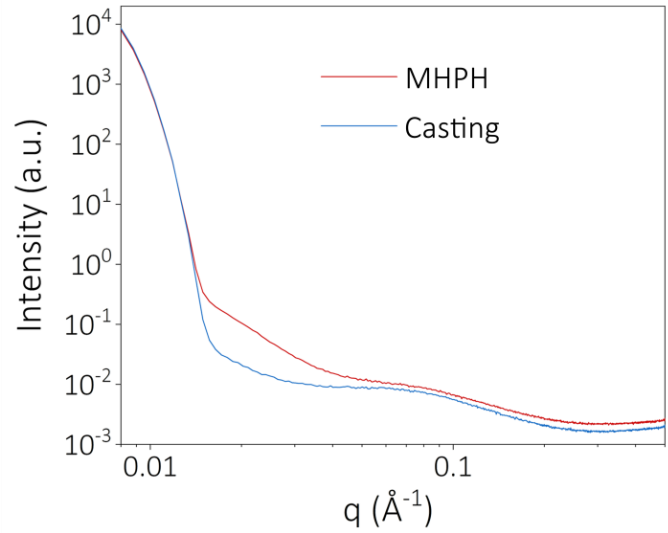

**Fig S4.** SXAS curves of casting and MHPH hydrogel films (thickness 1 mm).

An increased q value in the range of 0.02 Å$^{-1}$ to 0.06 Å$^{-1}$ is observed in MHPH, indicative of phase separation (main network separated from the highly entangled regions). We note the phase separation is beneficial for the realization of enhanced modulus, boosted toughness and low hysteresis hydrogels.

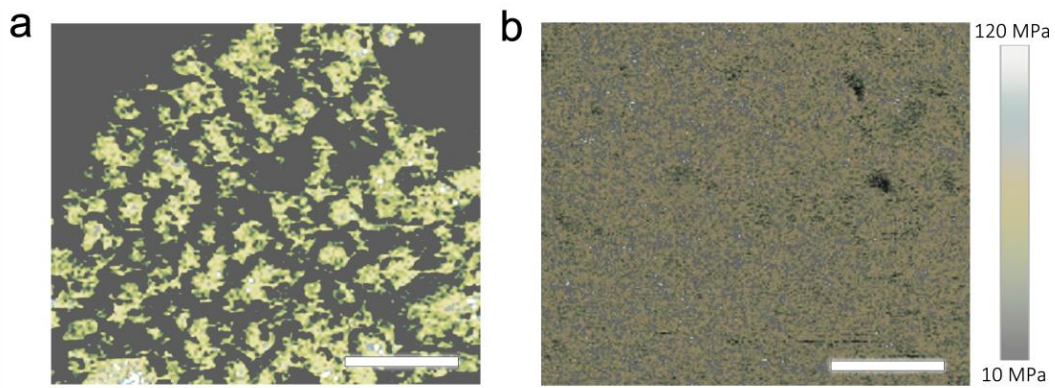

**Fig S5.** Modulus mapping of hydrogel films prepared by **a**, MHPH method and **b**, traditional casting hydrogel.

This figure set further confirms the phase separation in the MHPH hydrogel, in sharp contrast to the traditional casting hydrogel.

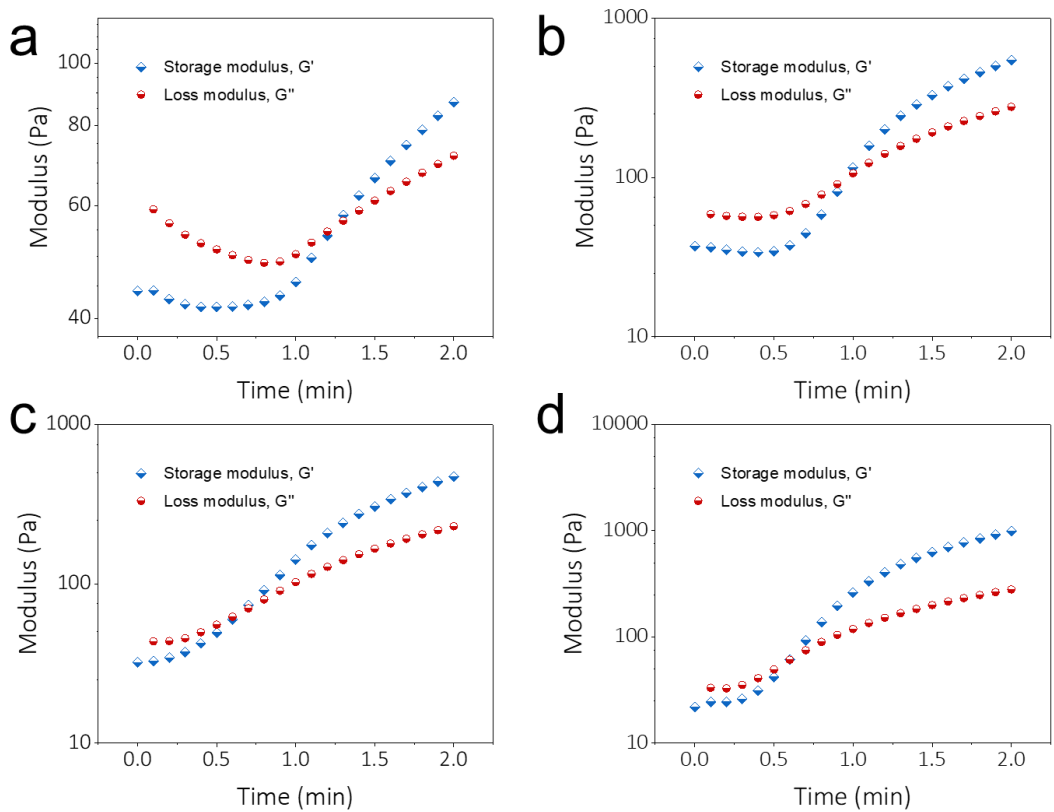

**Fig. S6**. The storage modulus and loss modulus of the ink with MXene concentration varying from (**a**) 1.0 mg mL$^{-1}$, (**b**) 3.2 mg mL$^{-1}$, (**c**) 8.5 mg mL$^{-1}$ to (**d**) 10.6 mg mL$^{-1}$, respectively. The curves were obtained by simultaneously injecting 20 μL of APS solution into the ink and began testing.

The ink storage modulus (G') and loss modulus (G'') can we easily adjusted by varying MXene concentration. When the ink G'' value is higher than the G', the gelation is still ongoing and the ink remains fluidic. However, when the G' surpasses G'', the ink is no longer flowable. Instead, the ink polymerizes into elastic solids and starts to behave as solid properties. The cross-point between G' and G'' curves indicates the solidification time.

The solidification time is MXene concentration-depend. Increasing the MXene content decreases the solidification time from 1.3 min (1.0 mg mL$^{-1}$, a), to 1.0 min (3.2 mg mL$^{-1}$, b), 0.7 min (8.5 mg mL$^{-1}$, c) and 0.6 min (10.6 mg mL$^{-1}$, d), respectively.

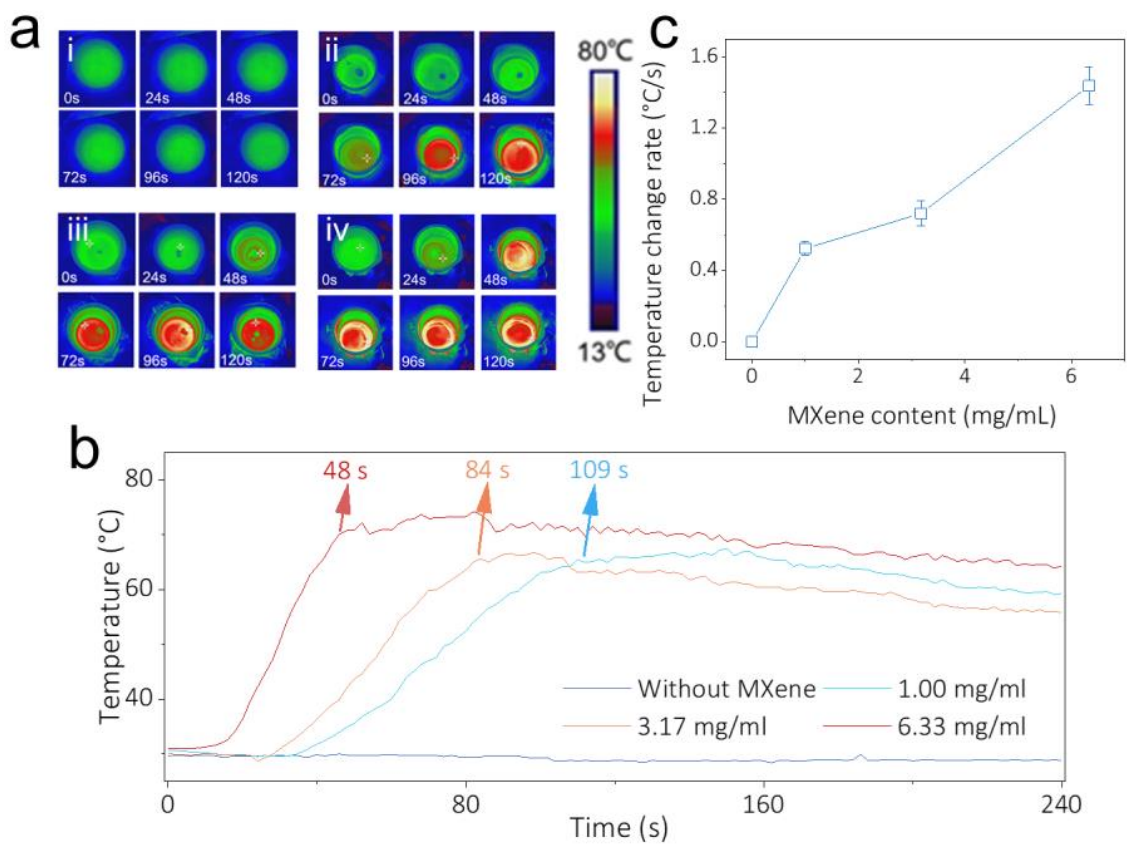

**Fig. S7. a**, The time-dependent infrared thermography with different MXene content. **b**, the corresponding temperature-time curves and **c**, The temperature change rate during the ink polymerization with different MXene contents.

Since polymerization is a self-exothermic reaction (Fig. S8a), the polymerization rate can be represented by the slope of the temperature time curve (Fig. S8b). A larger slope indicates a faster polymerization rate. We selected the highest temperature point as the reaction endpoint corresponding to the solidification threshold (Fig S8b). Fig S8c shows that the polymerization rate increases monotonically with MXene content, indicating that the polymerization rate can be accelerated by MXene.

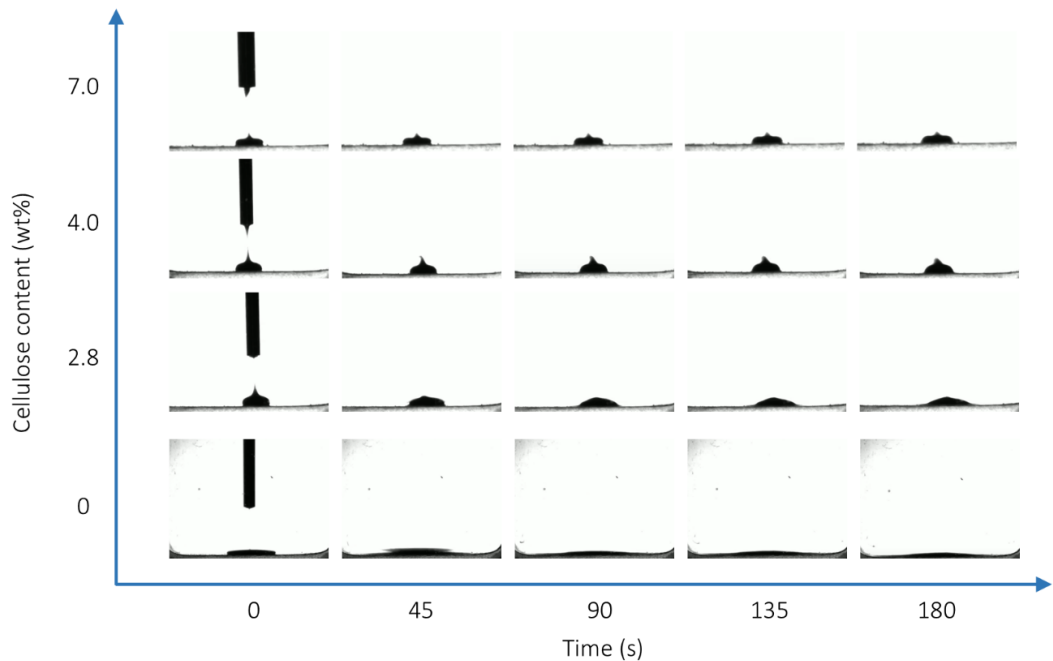

**Fig. S8**. Effect of cellulose content on the swelling of as-printed lines in water.

The addition of cellulose content largely controls the swelling of as-printed lines. When cellulose is absent, the as-extruded filament immediately decomposes in water. Increasing the cellulose content to 2.8 wt.% allows the printing of well-defined patterns, which quickly deforms due to the swelling in water. Further increasing the cellulose content to 4.0 wt.% leads to the well preservation of printed lines even after immersing in water for 180 s, demonstrating the critical anti-swelling capability of cellulose.

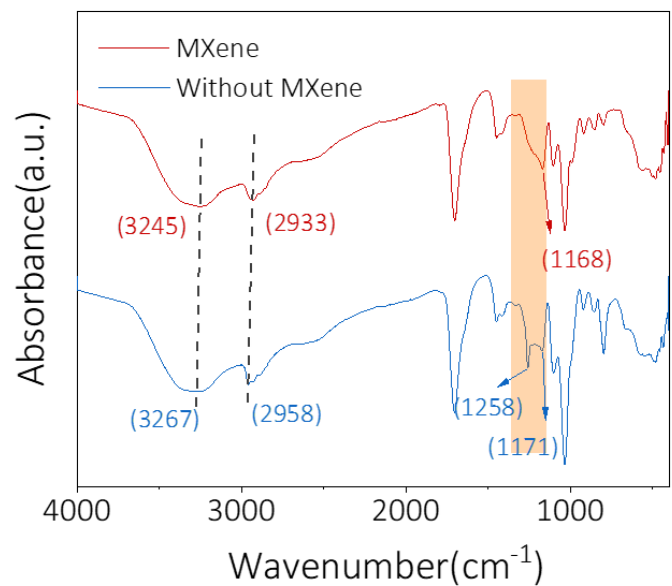

**Fig. S9**. The FT-IR spectrum of AAc monomer and thickener solution with or without MXene.

Incorporation of MXene into the original ink system (contain PAA, HPMC, MBAA and water) induced a redshift of the hydrogen bonding peak (2958 cm$^{-1}$) to lower wavenumbers, indicating the forming of a stronger hydrogen bonding interactions between hydroxyl groups on MXene surfaces and the hydroxyl/carboxyl moieties within the ink matrix.

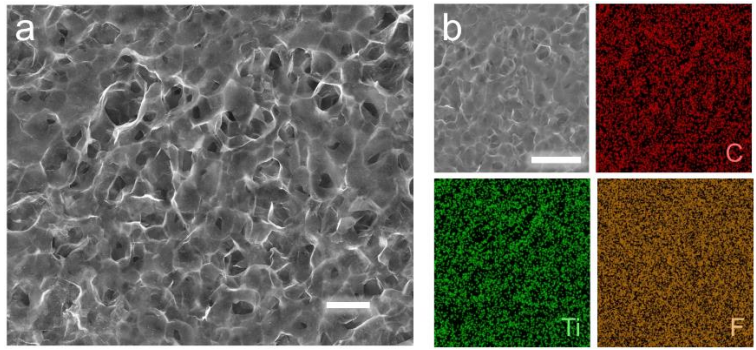

**Fig. S10. a**, The SEM image of the printed MHPH obtained through freeze-drying, the scale bar is 20 μm. **b**, The corresponding elements distribution, the scale bar is 40 μm. The as-printed MHPH exhibits a porous microstructure, with MXene uniformly distributed in the hydrogel matrix.

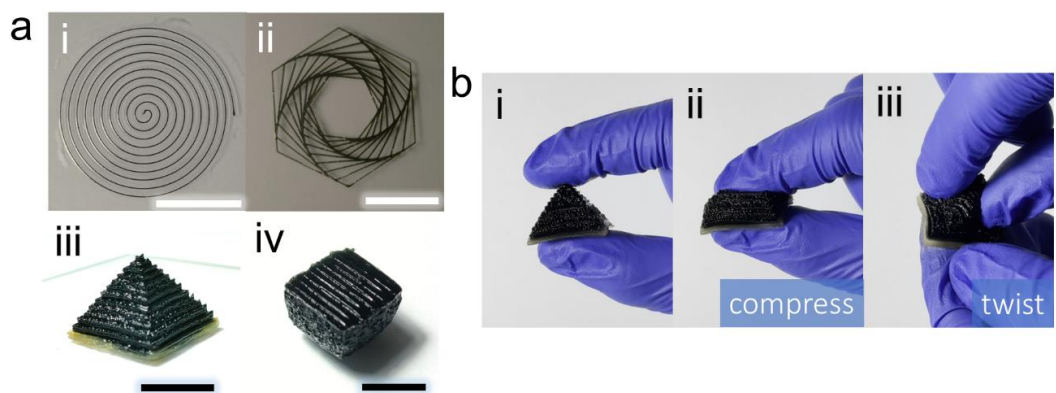

**Fig. S11. a**, 3D-printed planar patterns (**i**, **ii**) and 3D structures (**iii**, **iv**) of MHPH, the scale bar is 2 cm. **b**, The resilient nature of 3D-printed MHPH pyramid. Even after repeated pressing and twisting, the 3D-printed MHPH pyramid can quickly restore the original structure, fully demonstrating the robust nature of MXene-triggered 3D printed hydrogels.

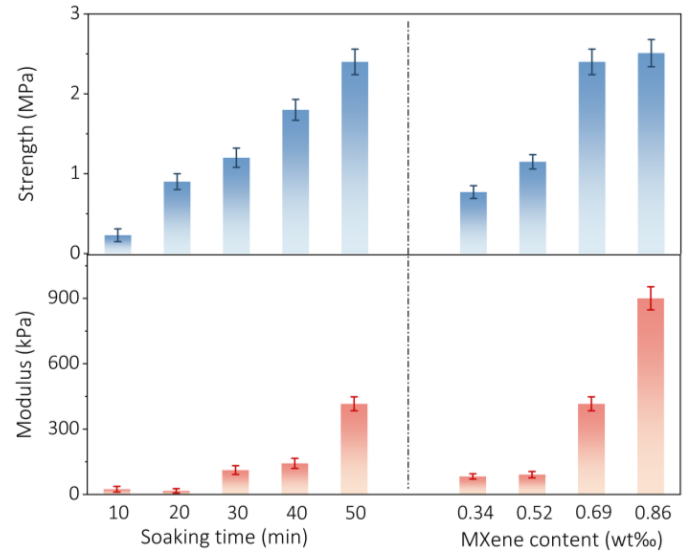

**Fig S12.** Effect of soaking time and MXene content on the hydrogels' modulus and strength.

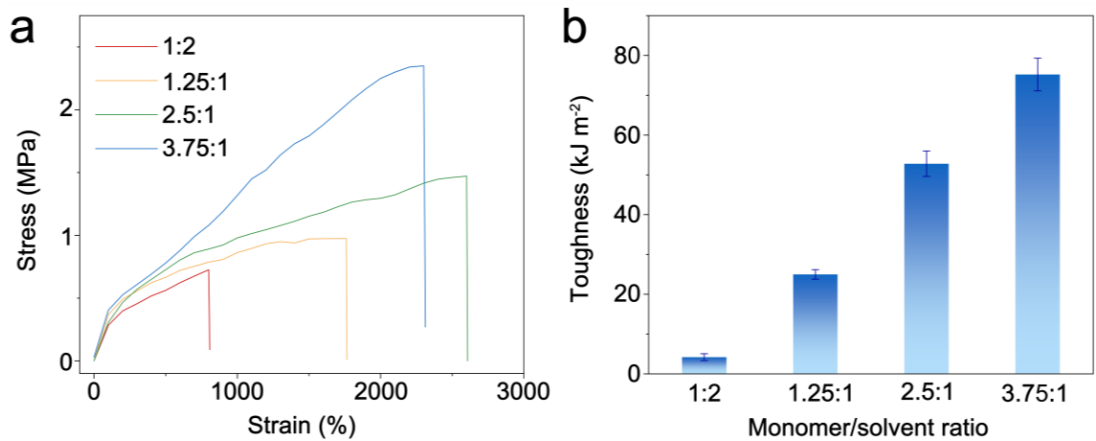

**Fig S13**. **a**, Tensile stress-strain curves and **b**, toughness statistics of hydrogels with different monomer/solvent molar ratio. As increasing the monomer content, more entangling regions are formed, leading to increased entangled density and substantially increased modulus and toughness.

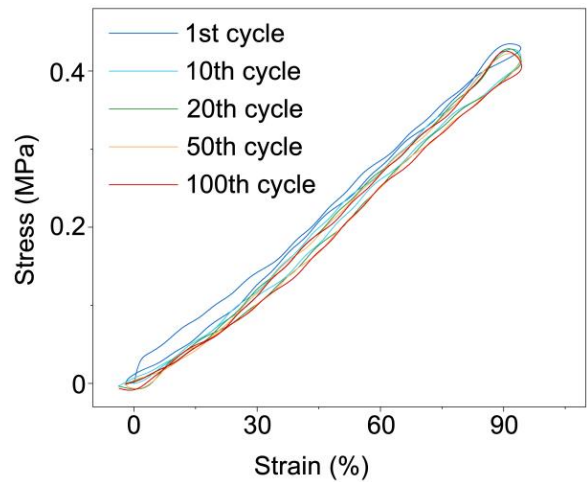

**Fig S14**. Hysteresis curves of hydrogel at different stretching cycles (soaking time is 50 min, MXene content is 0.69 wt‰ and the monomer/solvent molar ratio is 3.75:1). Clearly, the stress-strain curves can be well maintained without severely fluctuated even after repeated stretching to 100% for up to 100 cycles, fully demonstrating the stability of the polymerized network. The stable low hysteresis in the hydrogel can be fairly attributed to the phase separation between the hyperbranched regions and highly entangled regions.

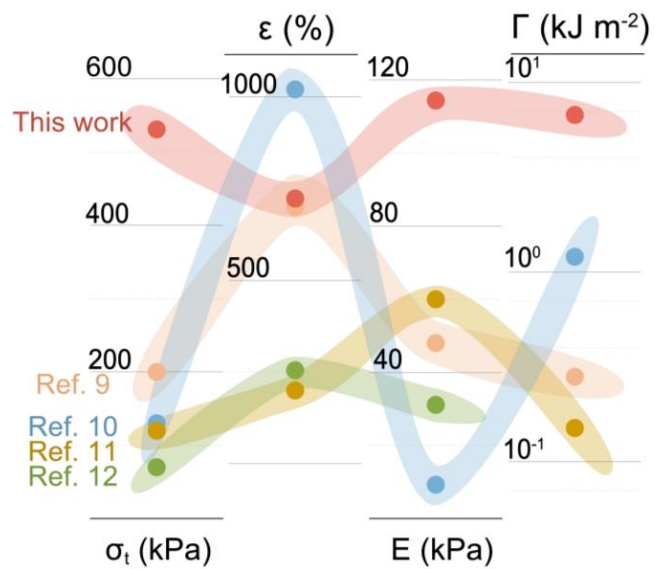

**Fig S15**. Comparison of key mechanical parameters of MHPH (based on polyacrylic acid as the monomer) with those reported in literature (*10-13*).

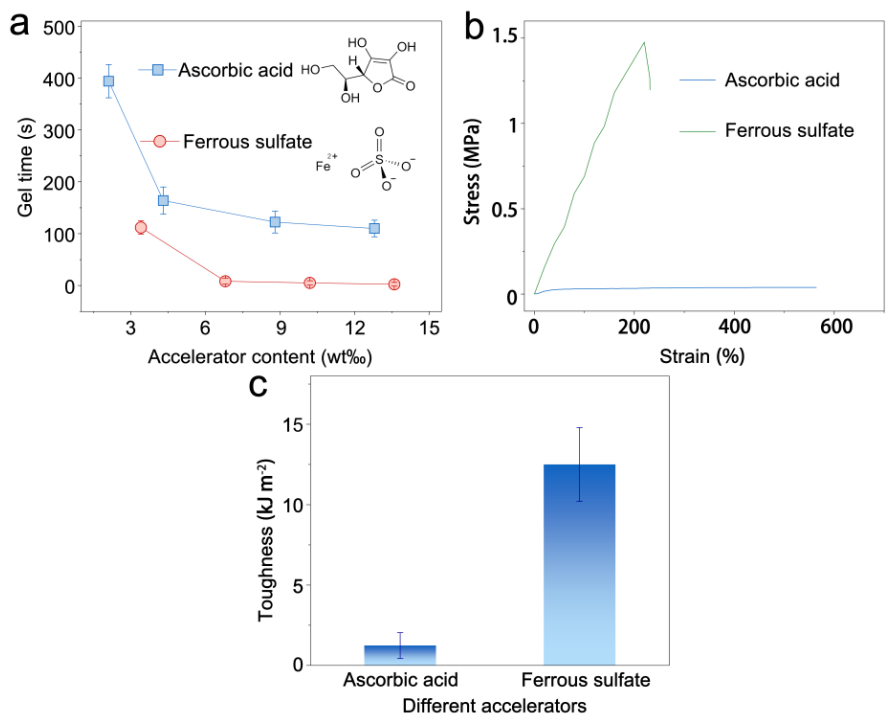

**Fig S16**. **a**, Effects of different accelerators and their content on the polymerization time (monomer is acrylic acid). **b** and **c**, tensile stress-strain curves and corresponding toughness comparison of hydrogels triggered by different accelerators.

Beside MXenes, other accelerators like ascorbic acid or ferrous sulfate also triggers the polymerization of acrylic acid. However, it takes much longer (two-order of magnitude more) to complete the polymerization compared to that of MXene, rendering the DIW impossible. The strength and toughness of the polymerized hydrogels triggered by ascorbic acid or ferrous sulfate are also much lower than those of MXene-triggered hydrogels.

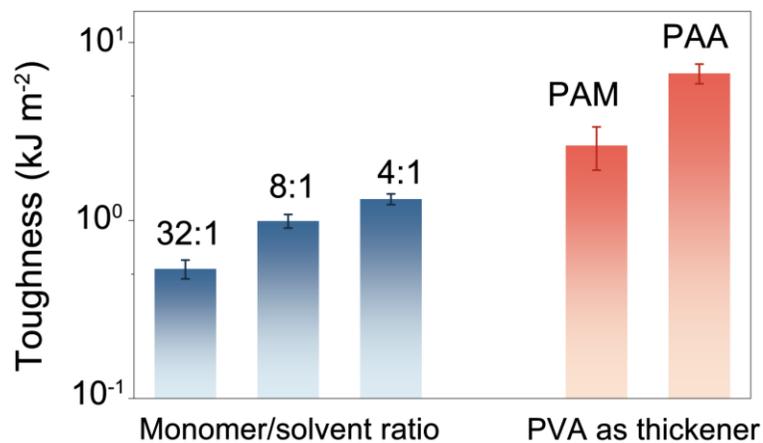

**Fig S17.** Toughness comparison of hydrogels with acrylamide as monomer and PVA as thickener

Increasing the amount of acrylamide monomer leads to the increased toughness in the resultant PAM hydrogels. Besides, by replacing the cellulose thickener with PVA, MXene can also rapidly trigger the polymerization using acrylic acid or acrylamide monomers, with a higher toughness realized in the PAA hydrogel. These results suggest that the MXene triggering chemistry is effective for different monomers or thickeners, indicating the generality of this novel heterogeneous polymerization route.

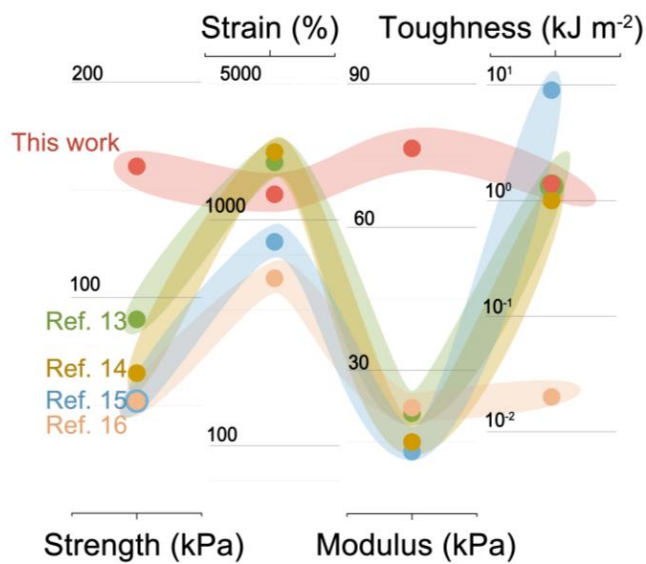

**Fig. S18.** Comparison of critical metrics of acrylamide-based hydrogels reported in literature (*14-17*) with polyacrylamide obtained through our heterogeneous polymerization strategy.

From this figure, our MHPH sample exhibits the highest strength and modulus, and comparable strain and toughness with even the best samples. We note these important metrics-strength, modulus, strain and toughness- are realized in one single MHPH hydrogel, which has never been achieved in previous reports. However, due to the lower solubility of AM in water, the limited upper ratio of monomer to solvent renders the formation of highly entangled regions impossible.

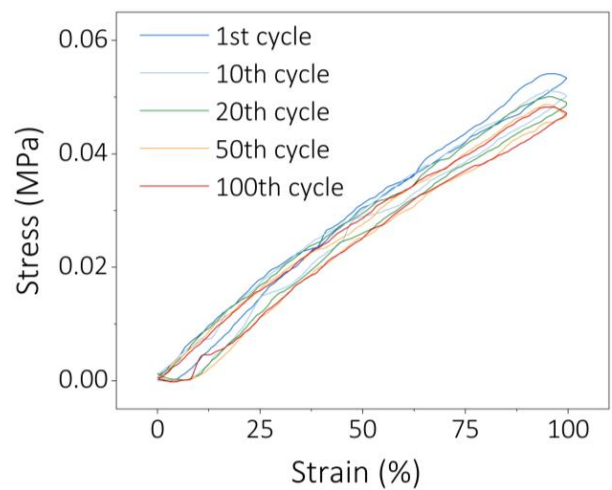

**Fig S19**. Hysteresis curves at different cycles of the PAM hydrogel (monomer/solvent molar ratio is 1:4), showing a relatively small hysteresis even after 100 cycles of stretching.

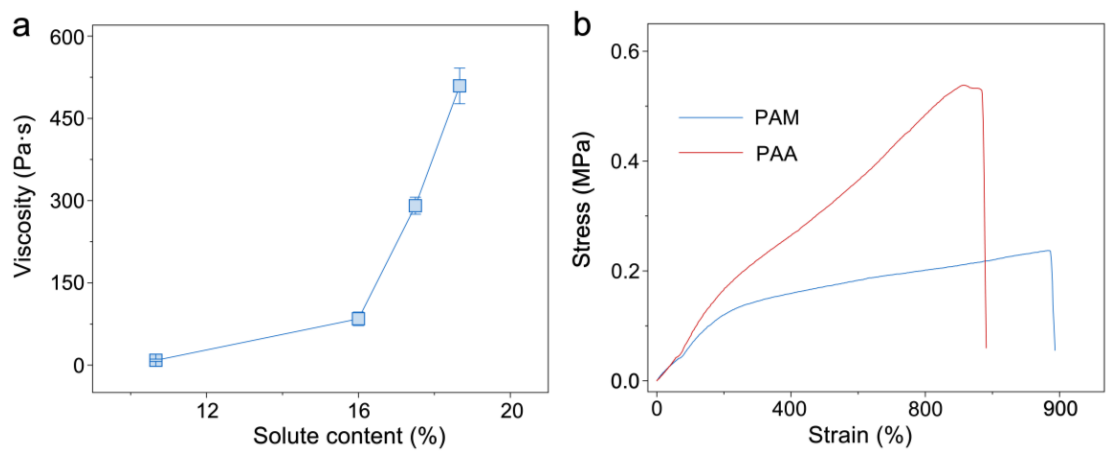

**Fig S20**. **a**, The relationship between polyvinyl alcohol (PVA) content and viscosity and **b**, The tensile curves of hydrogels with PVA as thickener.

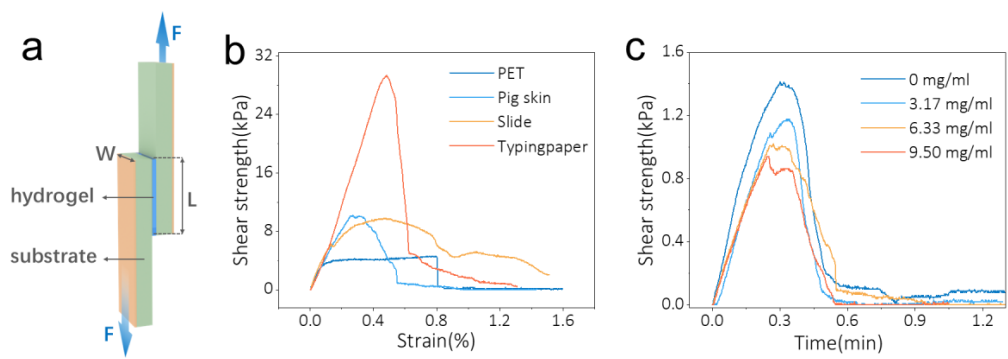

**Fig. S21**. The adhesion force of polyacrylic acid obtained through heterogeneous polymerization strategy using PVA as a thickener **a**, Schematic diagram of adhesion testing. **b** on different substrates and **c**, with different MXene contents on pig skin.

To quantitatively determine the adhesion force of MHPH on the substrate, we designed the test model as shown in Fig S21a, where the hydrogel is placed among two identical substrates. Since the adhesion force equals to the pulling force, then one is able to detect the adhesion force by recording the pulling force till the depart of MHPH from the substrate.

As shown in b, different substrates indeed affect the adhesion force, which reaches the highest in the typing paper and the lowest in the PET. At a given substrate (i.e., pig-skin), then MXene content also affects the adhesion force.

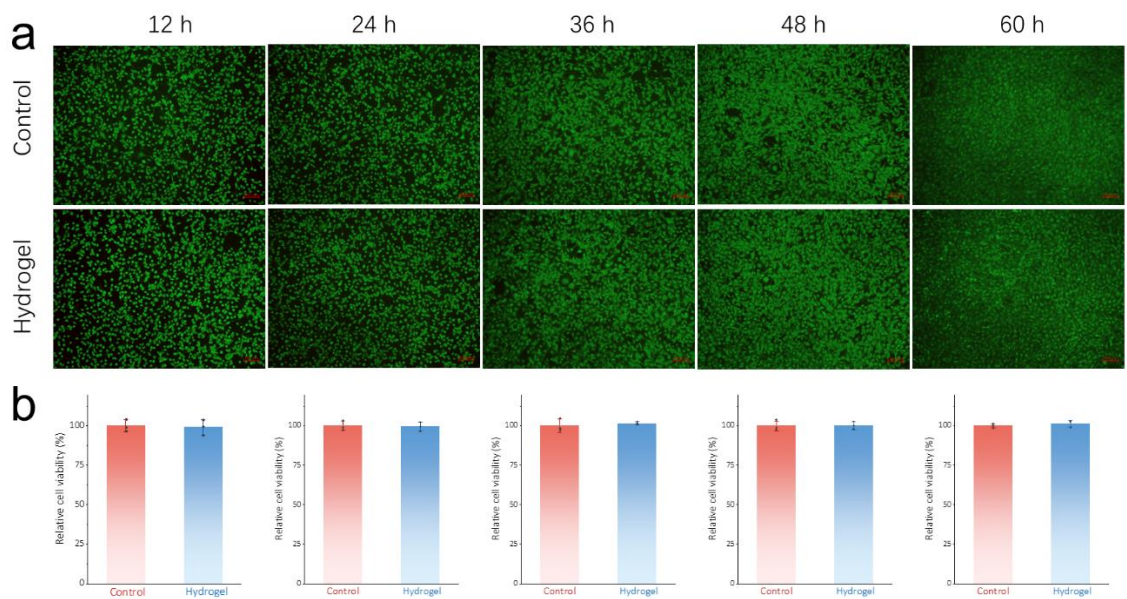

**Fig. S22. Cytotoxicity test.** a, Survival of NIH3T3 cells co cultured with samples for different time by Live/dead staining. b, Cell viability statistics of samples co cultured with NIH3T3 cells for different durations by CCK8 method. Based on the cytotoxicity test, no obvious toxic effects on cells are observed, indicating the biofriendly nature of the hydrogel.

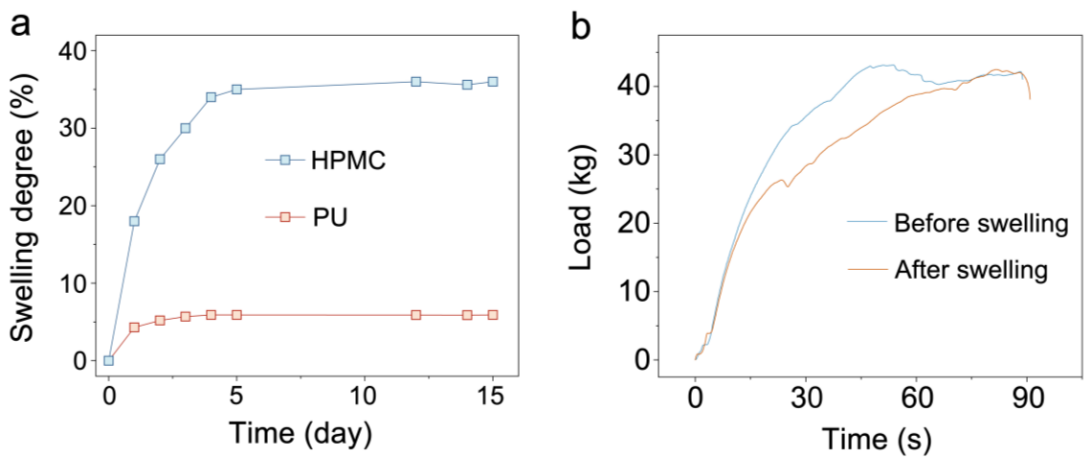

**Fig. S23**. Anti-swelling properties of MHPH using different thickeners. **a**, The anti-swelling performance curves of MHPH in water using polyurethane (PU) and HPMC as thickeners. **b**, The tensile stress-strain curves before and after soaking in water for 15 days with PU as a thickener.

Compared to the conventional hydrogel, TCH, MHPH shows a much-reduced swelling degree. Even after soaking in water for 15 days, the MHPH still exhibits a similar load as freshly-printed, demonstrating the excellent anti-swelling property in the MXene-triggered hydrogel.

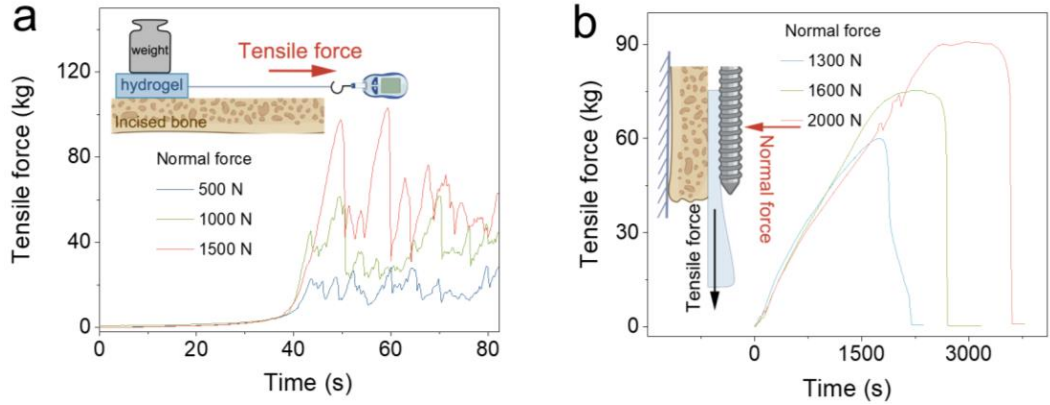

**Fig. S24. a**, Measurement of maximum static friction between hydrogel and bone tunnel. **b**, Tensile curves of hydrogel under different normal forces.

In order to meet the requirement of different tensile forces, we applied a normal force on the hydrogel by squeezing the screw and bone tunnel. **Fig. 24b** shows that the hydrogel is not damaged under the given normal force and can function properly.

Firstly, the coefficient of static friction can be calculated through the following formula:

$$f = \mu F \qquad (1)$$

where $f$ is the maximum static friction force, which is equal to the peak tensile force, $\mu$ is the coefficient of static friction and **F** is normal force. Here the calculated value of $\mu$ is **0.457** (**Fig. 24a**). When the normal forces are 1300 N, 1600 N and 2000 N, and the screw connection part can withstand 60 kg, 75 kg and 90 kg without slipping and breakage, respectively (**Fig. 24b**). This demonstrates that MHPH is compatible with screw-based fixation techniques in ligament repair surgery.

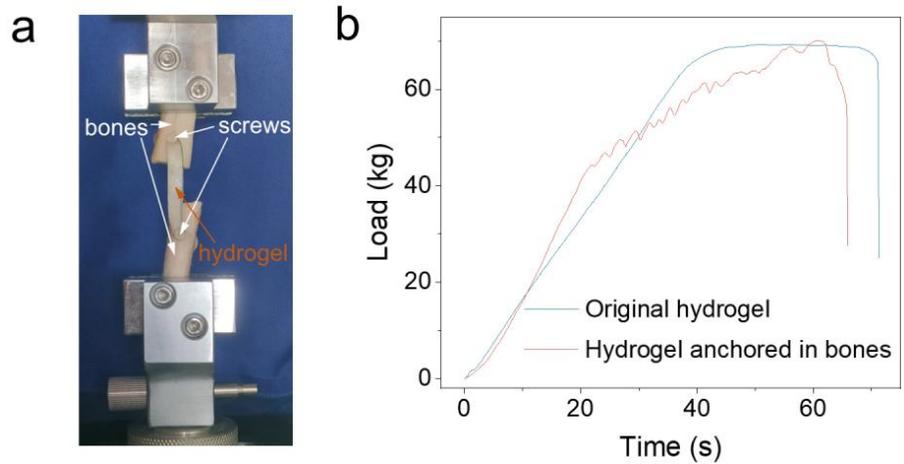

**Fig. S25. a**, Real image of the MHPH hydrogel anchored in bones with degradable screws. **b**, The mechanical measurements of MHPH before and after being anchored in bones, showing no much mechanical differences. This lays a solid basis for the potential artificial ligament repairment.

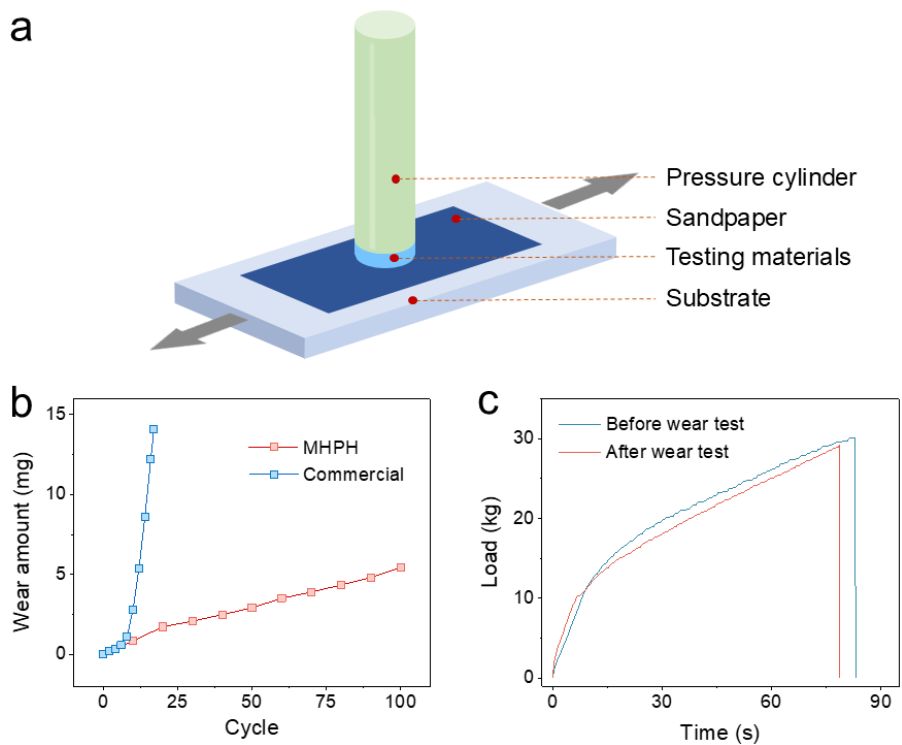

**Fig. S26**. **Wear resistance performance**. **a**, Schematic diagram of homemade instrument for wear resistance testing. **b**, The wear performance comparison between MHPH and commercial artificial ligaments LARS. **c**, Tensile mechanical curves of MHPH before and after 100 cycles of wearing.

    Importantly, our MHPH exhibits excellent anti-wearing performance which is much superior to the commercial one, LARS (Fig S25b). Even after 100 cycles of wearing, the MHPH still demonstrates well-preserved mechanical curve, fully indicating the excellent anti-wearing properties (Fig S25c).

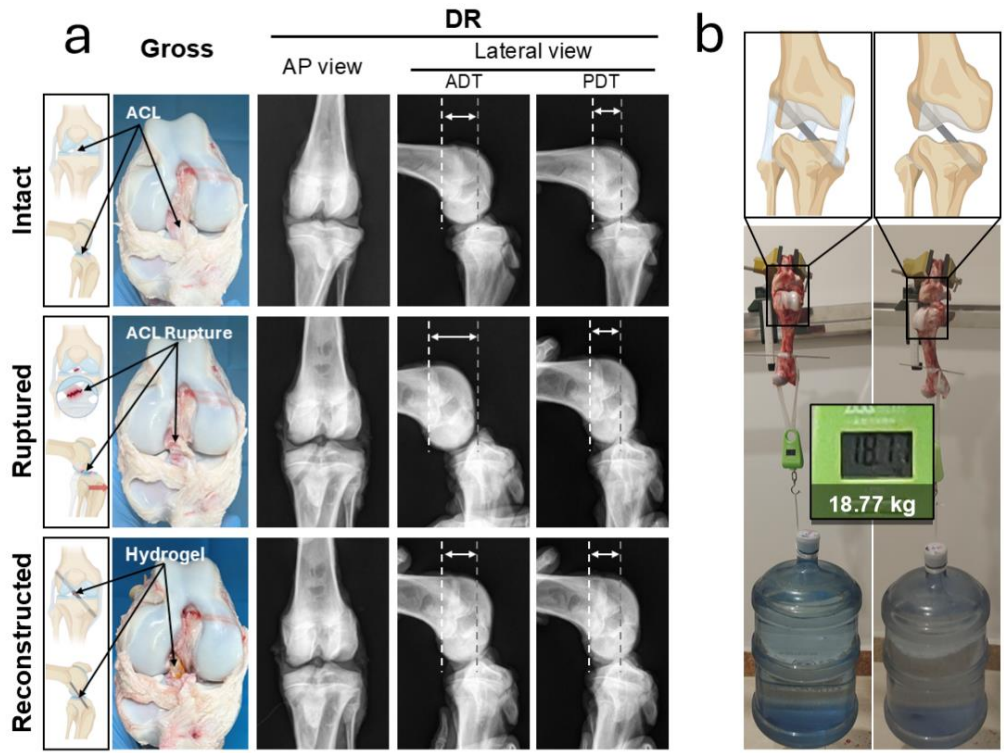

**Fig. S27 a**, The gross pictures of reconstruction and digital radiography (DR) of the anterior drawer test. **b**, Photos of the reconstructed knee joint lifting heavy objects.

The anterior cruciate ligament (ACL) is one of the dominant ligaments involved in maintaining knee stability and is the most frequently injured ligament in the knee. The tibia will shift forward relative to the femur under the stress of anterior drawer force, when the ACL is ruptured. ACL reconstruction using the MHPH could successfully restore the knee stability. The reconstructed knee could easily withstand a heavy object (total weight of 18.77 kg) suspended from the femur (Fig S26b, left panel), even when all knee ligaments and soft tissues were removed except the reconstructed ACL (Fig S26b, right panel).

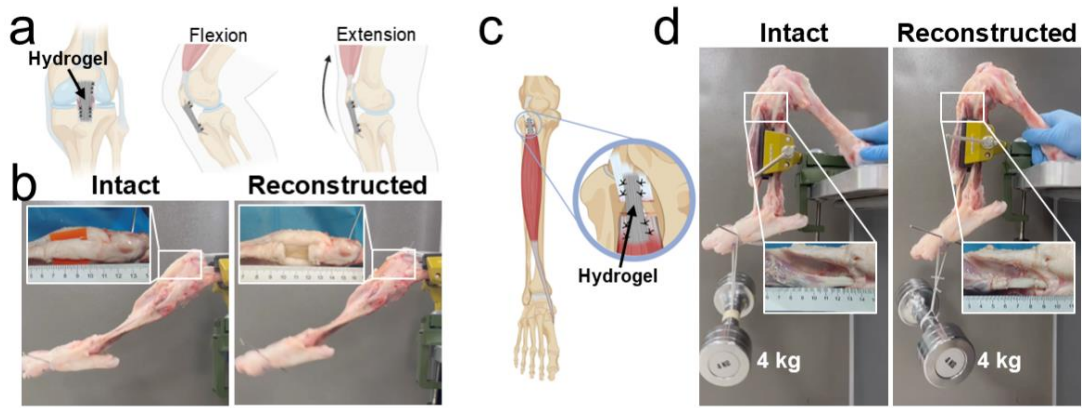

**Fig. S28**. **a**, Schematic diagrams and **b**, actual item photos of MHPH used for patellar ligament (PL) reconstruction. **c**, Schematic diagrams and **d**, actual item photos of MHPH used for anterior tibial tendon reconstruction.

Tendons and ligaments are not only important for the stability of joints, but also for their mobility. For example, PL plays a critical role for knee extension and anterior tibial tendon is important for ankle dorsiflexion. Tendon or ligament reconstruction using the MHPH could successfully restore the corresponding joint mobility.

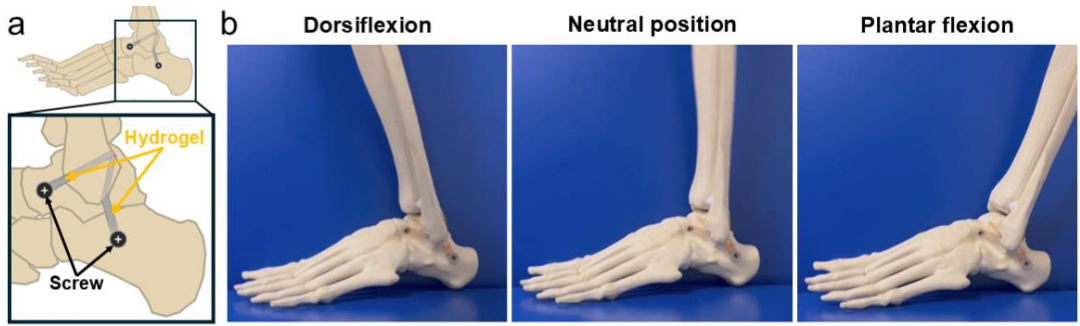

**Fig. S29**. **a**, Schematic diagram of anterior talofibular ligament (ATFL) and calcaneofibular ligament (CFL) reconstruction surgery. **b**, The ankle activity after ATFL and CFL reconstruction by MHPH in an ankle model.

For artificial tendon/ligament grafts, sufficient flexibility is required to avoid affecting joint movement after reconstruction. Even after being folded during complex ligament reconstruction surgery, our MHPH can still maintain sufficient flexibility and will not interfere with the normal movement of the joint after implantation.

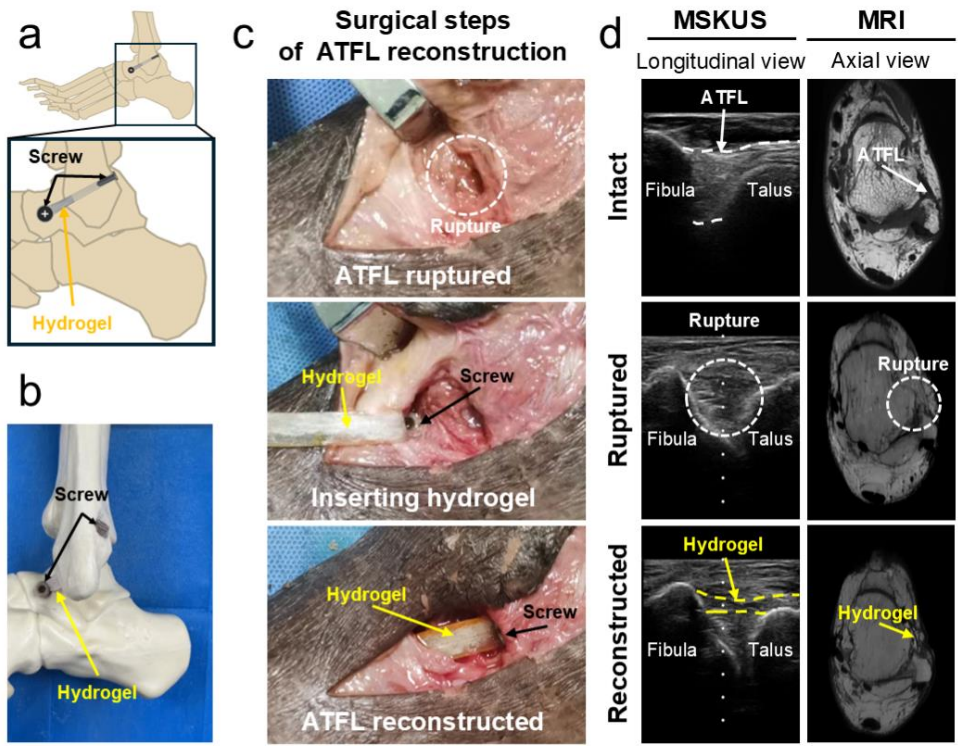

**Fig. S30**. **a**, Schematic diagram of anterior talofibular ligament (ATFL) reconstruction surgery. **b**, The actual demonstrative item of ATFL reconstruction by MHPH in an ankle model. **c**, The actual demonstrative item of ATFL reconstruction surgical steps by MHPH in a human sample. **d**, The musculoskeletal ultrasound (MSKUS) and magnetic resonance imaging (MRI) of the reconstructed ATFL in a human sample.

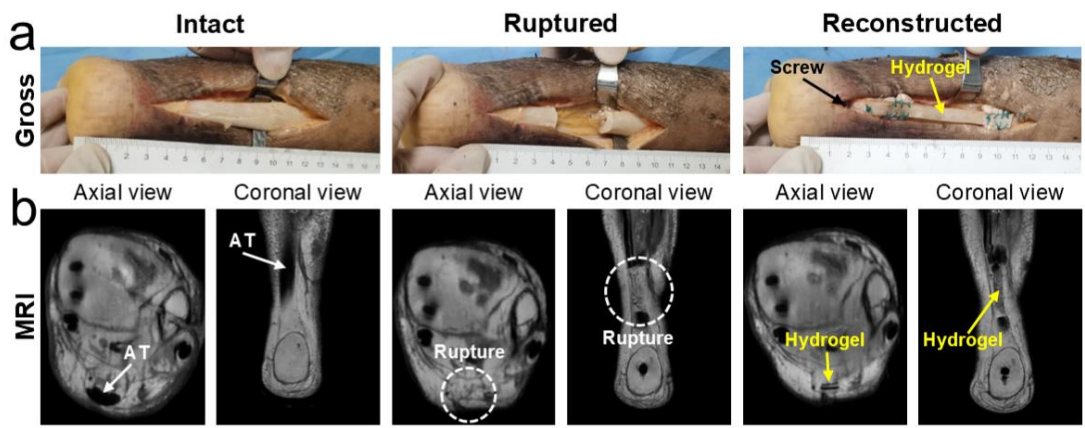

**Fig. S31. a**, Actual item photos and **b**, magnetic resonance imaging (MRI) of MHPH used for Achilles tendon reconstruction in a human sample.

In addition to excellent mechanical properties and good biocompatibility, our MHPH also meets the requirements of fixation methods, surgical manipulations, and radiological detections, which are commonly used in the clinic for ligament/tendon reconstruction.

| Strain | Strength (MPa) | Toughness (MJ m$^{-3}$) | Hysteresis | Topological structure | 3D Print (Y/N) | Ref. |
|---|---|---|---|---|---|---|
| 10 | 0.86 | 3.38 | 0.425 | Dual-network | Y | *Adv. Funct. Mater.* **2021**, 22, 2107202. |
| 2 | 0.9 | 0.125 | 0.33 | Dual-network | Y | *Adv. Funct. Mater.* **2021**, 31, 2100462. |
| 3.5 | 0.39 | 0.6 | 0.01 | High entanglement | N | *Science* **2021**, 374, 212. |
| 6 | 0.5 | 0.85 | 0.01 | High entanglement | N | *Adv. Mater.* **2022**, 34, 2206577. |
| 4.4 | 0.22 | 0.34 | 0.02 | High entanglement | N | *Adv. Mater.* **2023**, 35, 2210021. |
| 12 | 5.5 | 22 | 0.02 | Slide-ring | N | *Science* **2021**, 372, 1078. |
| 9 | 94.6 | 310 | 0.4 | Dual-network | Y | *Nature* **2024**, 631, 783. |
| 4 | 0.22 | 8 | 0.126 | High entanglement | Y | *Science* **2024**, 385, 566. |
| 15.7 | 14.8 | 83.6 | 0.38 | High entanglement | Y | *Adv. Mater.* **2023**, 35, 2304430. |
| 26 | 2.5 | 24.6 | 0.029 | Heterogeneous structure | Y | This work |

**Table S1**. Comparison of MHPH with previously reported hydrogels.